\documentclass[11pt]{article}
\pdfoutput=1  % togliere il commento se le figure sono pdf, Blasciarlo se sono eps
\usepackage{jcappub}
\usepackage{graphicx}
\usepackage{amsmath,amssymb}
\graphicspath{{./figures/}}
\usepackage{appendix}
\usepackage{placeins}
\usepackage[normalem]{ulem}
\usepackage[utf8]{inputenc}
\usepackage{braket}
\usepackage{multirow}
\usepackage{float}
\usepackage[table,x11names]{xcolor}
\usepackage{comment}

\usepackage{mathtools}
\usepackage{amsfonts}
\usepackage{upgreek}
\usepackage{latexsym}
\usepackage{stfloats}
\usepackage{afterpage}

\newcommand{\be}{\begin{equation}}
\newcommand{\ee}{\end{equation}}
\newcommand{\beq}{\begin{equation}}
\newcommand{\eeq}{\end{equation}}
\newcommand{\bea}{\begin{eqnarray}}
\newcommand{\eea}{\end{eqnarray}}

\newcommand{\bx}{{\mathbf  x}}

\newcommand{\bn}{{\mathbf  n}}
\newcommand{\bk}{{\mathbf k}}

\newcommand{\bV}{{\mathbf V}}

\newcommand{\bnabla}{{\boldsymbol{\nabla}}}

\newcommand{\dd}{\partial}
\newcommand{\HH}{{\cal H}}

\newcommand{\De}{\Delta}

\newcommand{\RR}{{\cal R}}

\newcommand{\al}{\alpha}

\newcommand{\de}{\delta}

\newcommand{\La}{\Lambda}
\newcommand{\si}{\sigma}

\newcommand{\om}{\omega}
\newcommand{\Om}{\Omega}

\newcommand{\cd}{\cdot}
\newcommand{\ra}{\rightarrow}
\newcommand{\pa}{\parallel}

 \definecolor{magenta}{rgb}{0.1,0.98,0.6}

\definecolor{dgreen}{rgb}{0,0.6,0.0}
\newcommand{\rut}[1]{{\color{black}#1}}

\definecolor{cyan}{rgb}{0,0.6,0.6}
\newcommand{\wil}[1]{{\color{black}\tt #1}}

\definecolor{red}{rgb}{0.8,0.2,0.2}
\newcommand{\den}[1]{{\color{black}\tt #1}}

\newcommand{\ed}[1]{{\color{black} #1}}

\newcommand{\CC}{\cellcolor{lightgray}}

\title{Redshift weighted galaxy number counts}
\author{{William L. Matthewson},}
\author{{Dennis Stock}}
\author{and {Ruth Durrer}}
\affiliation{Universit\'e de Gen\`eve, D\'epartement de Physique Th\'eorique and Centre for Astroparticle Physics,
24 quai Ernest-Ansermet, CH-1211 Gen\`eve 4, Switzerland}

\emailAdd{william.matthewson@unige.ch}
\emailAdd{dennis.stock@unige.ch}
\emailAdd{ruth.durrer@unige.ch}

\abstract{In this paper we introduce the `redshift fluctuation' as a gauge-invariant cosmological observable and give its fully relativistic expression at first order in cosmological perturbation theory. \rut{We show that this corresponds effectively to number counts with a radial window function with vanishing mean which therefore resolve smaller scale radial modes than standard number counts.} In a detailed analysis of the angular power spectrum of this new variable, we study the relevance of different relativistic contributions, and how it differs from the conventional observable galaxy number count fluctuations. In order to investigate its utility for future spectroscopic surveys, we perform Fisher forecasts for a Euclid-like and an SKAII-like configuration, as examples. Particular focus is placed on the dependence of the results on the size of the redshift bins and on the cutoff in $\ell$ adopted in the analysis.
\vspace{0.5cm}\\  \centerline\today}

\begin{document}

\maketitle

\section{Introduction}
Recently, measuring the `redshift fluctuations' of galaxies has been advocated as a new cosmological observable and its power spectrum, as well as cosmological constraints which can be derived from this observable, have been studied~\cite{Hernandez-Monteagudo:2019epd,Hernandez-Monteagudo:2020xnl,Legrand:2020sek}. Here we follow up on this work; we derive and study a relativistic perturbative expression for `redshift fluctuations'. Some care is needed  when talking about `redshift fluctuations'. Of course, redshift is an observable, but its background value cannot be observed and therefore neither can its fluctuation.
It is well known that redshift fluctuations are gauge-dependent, we can e.g. choose our time variable to be a function of the observed redshift which implies that on these surfaces of constant time redshift fluctuations vanish.

Nevertheless, the method proposed in Refs.~\cite{Hernandez-Monteagudo:2019epd,Hernandez-Monteagudo:2020xnl,Legrand:2020sek}, to measure the redshift of each galaxy inside a redshift window defined by the observer and inside a small angular patch and to correlate these measures from different angular patches in the same or different redshift windows, is a well-defined observational procedure, leading to well-defined angular correlation functions  or power spectra.

In this paper we derive and investigate this `redshift fluctuation'  variable in a fully relativistic way at first order in perturbation theory. As an observable, it is of course gauge invariant. We study the relativistic contributions to the angular power spectrum of the redshift fluctuations and discuss its interest as a new cosmological observable.
We also generate Fisher forecasts for cosmological parameters measured with this new observable and with galaxy number counts. \rut{After we had submitted this paper a similar study, which however concentrates on cross correlations of the redshift weighted galaxy number counts with CMB data has appeared \cite{Lima-Hernandez:2022twl}. }

The paper is organized as follows. In the next section we briefly discuss the gauge issue which has been entirely neglected in previous papers on the topic. In Section~\ref{s:zeta} we introduce the gauge-invariant `redshift-weighted power spectrum' and show that it is identical to what was called `redshift fluctuation' in previous literature. \rut{We then move on to test the constraining power of the new variable in combination with galaxy number counts for estimating cosmological parameters. We make Fisher forecasts for future spectroscopic surveys similar to Euclid and to SKAII, whose specifications we provide in Section~\ref{s:fish}.} In Section~\ref{s:res} we discuss our numerical results and in Section~\ref{s:con} we conclude.
Some important formulae and derivations which can also be found elsewhere in the literature have been collected in several appendices for convenience.

{\bf Notation}~  We consider perturbations on a spatially-flat Friedmann universe with background metric $ds^2 = a^2(\tau)[-d\tau^2 +\de_{ij}dx^idx^j]$. Here $a$ is the scale factor, $\tau$ is conformal time and we shall often use the conformal Hubble parameter, $\HH \equiv\dot a/a \equiv(da/d\tau)/a$. It is related to the physical Hubble parameter by $H=\HH/a$. We normalize the scale factor to 1 today, so that it is related to the background redshift  by $a=1/(z+1)$.

\section{The gauge issue}
\rut{We consider a Friedmann metric with the most general scalar perturbations. It is  of the form}
\be
ds^2 = a^2\left\{-(1+2A)d\tau^2 +B_id\tau dx^i +\left[(1+2H_L)\de_{ij}+H_{T\, ij}\right]dx^idx^j\right\} \,,
\ee
where 
$$
B_i= \dd_iB ~~\mbox{ and }~~H_{T\, ij} = \left(\dd_i\dd_j -\frac{\de_{ij}}{3}\De\right)H_T 
$$
are derived from the scalars $B$ and $H_T$, which are related in a non-local way to the metric perturbations.
\rut{The four functions $A,~B$, $H_L,~H_T$ fully describe the perturbed metric and we consider them at first order only.}
The linear perturbation of the redshift, $\de z$ in an arbitrary gauge, for scalar perturbations  has been calculated, e.g., in~\cite{Bonvin:2011bg} with the result
\be\label{e:dzgauge}
\de z =-(1+z)\left[ H_L -\frac{1}{3}\De H_T +\bn\cdot\bV +\Phi+\Psi + \int_0^{r(z)}(\dot\Phi+\dot\Psi)dr\right] \,. 
\ee
where $r(z)$ is the comoving distance to redshift $z$.  For  vanishing spatial curvature, $K=0$,
\be
r(z) = \int_0^z\frac{dz'}{H(z')} \,.
\ee
 The gauge invariant variables $\bV$, the velocity on longitudinal gauge, as well as the Bardeen potentials $\Phi$ and $\Psi$, are best given in Fourier space where they are defined by, see e.g.~\cite{Durrer:2020fza}\footnote{In~\cite{Durrer:2020fza}, $H_T\ra k^2H_T$ and $B\ra kB$ so that the perturbation variables are dimensionless in real space. With our definition $B$ has units of length and $H_T$ has units of (length)$^2$.},
\bea
\bV &=& i\hat\bk V  \qquad   V= v-k\dot H_T\\
%\si = k(\dot H_T-B)\\
\Phi &=& -H_L -\frac{k^2}{3}H_T  +\HH(\dot H_T-B)\\
\Psi &=& A -\HH (\dot H_T-B) -(\ddot H_T-\dot B) \,.
\eea
Here $\hat\bk = \bk/k$ and  $i\hat\bk v$ is the peculiar velocity of the cosmological fluid in the (arbitrary) gauge under consideration in Fourier space. It is well known that $\Phi$, $\Psi$ and $\bV$ are gauge invariant, but $H_L$ and $H_T$ are not.
In longitudinal gauge, where $H_T=B=0$, we have $\Phi=-H_L$ so that \eqref{e:dzgauge} reduces to
\be\label{e:dzlong}
\left(\de z\right)^{\rm long} =-(1+z)\left[\bn\cdot\bV +\Psi + \int_0^{r(z)}(\dot\Phi+\dot\Psi)dr\right] \,. 
\ee
The first term, the so called Doppler term, of this expression is used in Refs.~\cite{Hernandez-Monteagudo:2019epd,Hernandez-Monteagudo:2020xnl,Legrand:2020sek} for the redshift perturbation $\de z$. The second and third term are the redshift contributions due to the gravitational potential and the integrated Sachs-Wolfe term\footnote{The authors of these Refs. mention also the second and third terms, but they neglect them as they are much smaller than the Doppler term.}.
This is the `Newtonian version', giving the change of redshift due to peculiar velocity and to the gravitational potential.
However, in practice we can only observe the full redshift $z=z_{\rm bg}+\de z$ where $z_{\rm bg}$ denotes the redshift in the background Friedmann universe. The split into a background and a perturbation is not unique.
It is easy to verify that under a (linearized) change of the time coordinate $\tau\ra \tau+T(\bx,t)$, the (linearized) redshift perturbation changes as
\be
\de z \ra \de z-(1+z)\HH T \,.
\ee
Therefore, from an arbitrary gauge with redshift perturbation $\de z$, choosing $T= \de z/[\HH(1+z)]$, we arrive at a new gauge with vanishing redshift perturbation. This renders the above expression dubious and actually non-measurable. In the next section we discuss what we can truly measure when assigning its redshift to each source .

\section{The redshift-weighted number count power spectra}\label{s:zeta}
We consider a survey with a given observed source density in redshift and angular space. These are the coordinates on the past lightcone to which the observer has truly access.

We denote by $N(z,\bn)dzd\Om_\bn$ the number of galaxies with redshift in the bin $[z,z+dz]$ and direction in the  solid angle $d\Om_{\bn}$ around the observation direction $\bn$.
Denoting by $\bar N(z)dz$ its mean over its angular extent, it has been shown in~\cite{Bonvin:2011bg} that
\be
\frac{N(z,\bn)-\bar N(z)}{\bar N(z)} \equiv \De(\bn,z) = \de_z(\bn,z) +\rut{\frac{\de v(\bn,z)}{v(z)}}
\ee 
is gauge invariant. Here, $\de_z=(\rho(z,\bn)-\bar\rho(z))/\bar\rho(z)$ is the \rut{matter} density fluctuation at fixed redshift $z$, while \rut{$v$ and $\de v$ are respectively the volume density  at fixed redshift and its fluctuation.
The galaxy overdensity at fixed redshift including linear bias is given in Appendix~\ref{a:pert}.} It has actually been found in~\cite{Bonvin:2011bg} that both $\de_z$ and $\de v$ are gauge invariant by themselves.
\rut{Overbars denote angular averages. We shall below identify them with background quantities which is in principle not completely accurate, since angular averages include 
perturbations of the size of the sphere considered. However, as the lowest redshifts we shall consider are $z \sim 0.2$, corresponding to a diameter of more than 1000$h^{-1}$Mpc, we neglect this difference in the calculations that follow.}

The fluctuation $\De(\bn,z)$ has also been derived in~\cite{2009PhRvD..80h3514Y,Yoo:2010ni} and Ref.~\cite{Challinor:2011bk} has included the important fact that we typically do not see all sources in a given direction at fixed redshift, but only those which arrive at our position with a flux which is higher than the limiting performance of our instrument (i.e. magnification bias). For completeness, we give the expressions of these perturbation variables in Appendix~\ref{a:pert}.
By construction, $\int d\Om_n\De(\bn,z) = 0$.

Assuming all matter to be clustered in objects of a fixed mean mass (i.e. neglecting biasing), up to a constant pre-factor,  the number fluctuation of the observed sources is
given by
\be
N(z,\bn) = \bar\rho(z)\bar v(z)\left[1+ \De(\bn,z) \right] \,, \quad \bar N(z) = \bar\rho(z)\bar v(z)\,.
\ee
Here $\bar \rho$ is the background matter density at redshift $z$ and $\bar v(z)$ is the  background `volume density' 
in redshift space. More precisely, in the background cosmology the infinitesimal volume at redshift $z$ in direction $\bn$ is given by
\be
dV  = \frac{r^2(z)}{(1+z)^3H(z)}dzd\Om = \frac{r^2(z)}{(1+z)^4\HH(z)}dzd\Om= \bar v(z)dzd\Om\,,
\ee
where $r(z)$ is the comoving distance to redshift $z$, see prior definition.
The matter density is simply $\bar\rho(z)=\rho_0(1+z)^3$.

We now consider the observed galaxies in some redshift bin, centred around some observer-defined redshift $z_1$, given by a normalized window function $W(z_1,z)$. We define the mean redshift of sources in this bin in direction $\bn$ by
\be\label{eq:zmean}
z_\bn(z_1) = \frac{\int dz\, z W(z_1,z)N(z,\bn)}{\int dzW(z_1,z)N(z,\bn)}\,.
\ee
The mean redshift of all the sources in our window around $z_1$ is given by
\be
\bar z(z_1) =  \frac{\int dz\, z W(z_1,z)\bar N(z)}{\int dzW(z_1,z)\bar N(z)}\,.
\ee
We can now define the  `redshift fluctuation' in the window as
\be
\zeta(\bn,z_1) := z_\bn(z_1)-\bar z(z_1) \,.
\ee
This quantity is truly observable. Its discrete version is the redshift fluctuation defined in Ref.~\cite{Hernandez-Monteagudo:2019epd}.
Introducing the modified window function
\be\label{e:deff}
\om(z_1,z) \equiv \frac{W(z_1,z)\frac{r^2(z)}{H(z)}}{\int dz\,W(z_1,z)\frac{r^2(z)}{H(z)} } \,,
\ee
we can rewrite $\zeta$ and $\bar z$ as
\be\label{e:exzeta}
\zeta(\bn,z_1)  =\int dz\, (z-\bar z(z_1))\De(z,\bn)\om(z_1,z)\,, \qquad
\bar z(z_1) =\int dz\, z\,\om(z_1,z)\,.
\ee
Note that by definition
$\int dz\, (z-\bar z(z_1))\om(z_1,z)=0$, hence the function $(z-\bar z(z_1))\om(z_1,z)$, which is entirely determined by the background, plays the role of a redshift weighting of the number count fluctuation $\De$, with vanishing mean.

\rut{The angular correlation function and the power spectrum of statistically isotropic functions on the sphere are, in full generality, related via the Legendre polynomials $P_\ell$. For the redshift fluctuation $\zeta$ at observer redshifts $z_1$ and $z_2$ this relation is

\begin{align}
    \langle \zeta(\bn,z_1)\zeta(\bn',z_2)\rangle &= \int dzdz'\om(z_1,z)\om(z_2,z')(z-\bar z(z_1))(z'-\bar z(z_2))\langle\De(z,\bn)\De(z',\bn')\rangle \label{e:Czeta1} \\
    &= \frac{1}{4\pi}\sum_\ell(2\ell+1)\hspace{-0.2cm}\int dzdz'\om(z_1,z)\om(z_2,z')(z-\bar z(z_1))(z'-\bar z(z_2))C_\ell^\De(z,z')P_\ell(\bn\cd\bn')\label{e:Czeta2} \\
    &\overset{!}{=}\frac{1}{4\pi}\sum_\ell(2\ell+1)C_\ell^\zeta(z_1,z_2)P_\ell(\bn\cd\bn')\quad\text{, thus}\\
    C_\ell^\zeta(z_1,z_2) &= \int dzdz'\om(z_1,z)\om(z_2,z')(z-\bar z(z_1))(z'-\bar z(z_2))C_\ell^\De(z,z')\quad.\label{e:Clzeta}
\end{align}
}

\rut{ For \eqref{e:Czeta2} we have used the relation between the correlation function and the power spectrum for the number counts variable $\De(z,\bn)$ and for (\ref{e:Clzeta}) we compare the coefficients of the Legendre polynomials, for more details see Appendix~\ref{a:pert}.

In the same way, one obtains}
\be
C_\ell^{\zeta\Delta}(z_1,z_2) = \int dzdz'\om(z_1,z)(z-\bar z(z_1))W(z_2,z')C_\ell^\De(z,z')\,.  \label{e:ClzetaD}
\ee

Like $\De$, these power spectra are gauge invariant and fully relativistic.
To determine $\zeta$, a good redshift resolution inside the considered window is required; hence, we need spectroscopic redshift data to measure $\zeta$. As we see from Eq.~\eqref{e:Clzeta}, the variable $\zeta$ contains the same relativistic contributions as the number counts $\De$.

In the following, we study  the angular power spectra of $\zeta$, $C_\ell^\zeta(z_1,z_2)$ and $C_\ell^{\zeta\De}(z_1,z_2)$, and investigate their dependence on the size of the redshift windows, as well as the relevance of the different contributions.

\subsection{Numerical examples}
In this section we study the angular power spectra of $\zeta$, based on a fully-relativistic description, in detail. We investigate the importance of the different contributions, the window function specifications, the correlation between different redshift bins, and the cross-correlation between galaxy number counts $\Delta$ and angular redshift fluctuations $\zeta$. All spectra are generated with a version of the CLASSgal code for galaxy number counts \cite{DiDio:2013bqa,CLASS}, modified to accommodate additionally the angular redshift fluctuations.\footnote{The modified version of \texttt{CLASS} along with the Fisher code used in this analysis may be found here:\\ \texttt{\url{https://github.com/WillMatt4/SMAL-FRY}}.}

The different contributions to the angular power spectrum given in equation (\ref{DezNF}) can be divided into four principal categories: density, redshift-space distortions, gravitational lensing, and large-scale gravitational effects, see appendix \ref{a:pert} for more details. Figure \ref{fig:contributions_euclid} shows the contributions of each effect to the total angular power spectrum. As one can see for both $\Delta$ and $\zeta$, the dominant contribution to the total power comes from the density, and on larger angular scales also from redshift-space distortions, whereas lensing and large-scale gravitational contributions are subdominant in both cases. The lensing contribution in $\zeta$ is even smaller than in the number counts. This is due to the fact that redshift-independent contributions vanish in $\zeta$ and the lensing term varies very slowly with redshift. The same is true for the large-scale gravitational contributions. The overall amplitude of $\zeta$ is at least four orders of magnitude lower than $\Delta$, essentially due to the  \rut{window function, which has vanishing mean: For $\zeta$, the fluctuation in the galaxy number counts is multiplied by the small deviation of the observed redshift from the mean in the window, cf. (\ref{e:exzeta}), which makes the $\zeta$-amplitude  significantly smaller. }

\begin{figure}[!ht]
    \centering
    \includegraphics[scale=0.7]{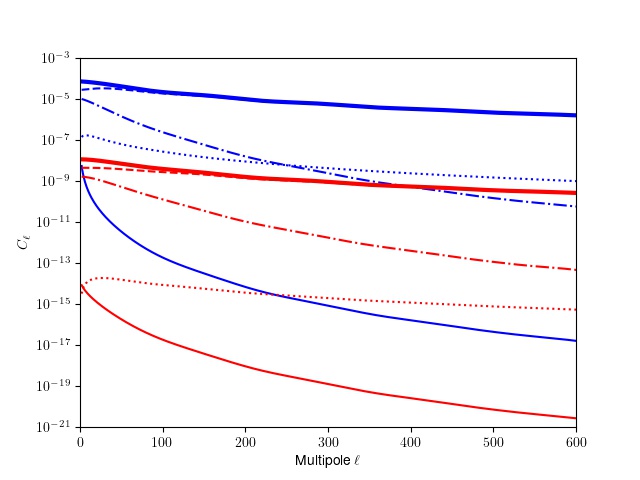}
    \caption{Typical angular power spectra for galaxy number counts $\Delta$ (blue) and angular redshift fluctuations $\zeta$ (red) based on a Euclid-like survey, cf. Table \ref{table:Specs}, for equal redshift correlations at $z=0.9225$ with bin widths $\si_z=0.02$. The total power is given by the thick solid line, the density contribution in dashed, the redshift-space distortions in dot-dashed, the lensing in dotted, and the large-scale gravitational terms by the thin solid line.}
    \label{fig:contributions_euclid}
\end{figure}
\begin{figure}[!ht]
    \centering
    \includegraphics[scale=0.7]{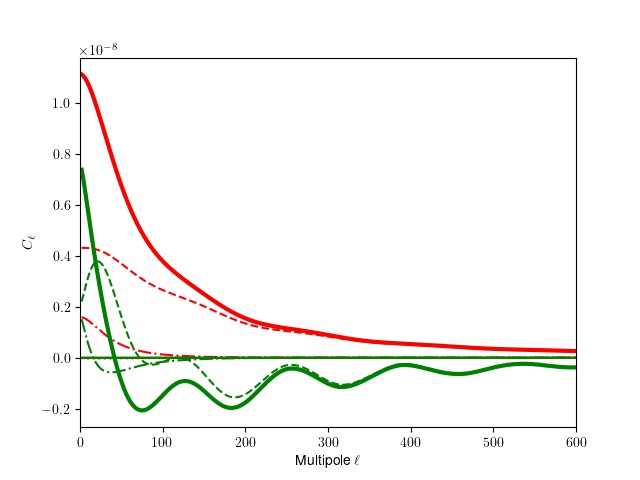}
    \caption{Typical angular power spectra $C_\ell^{\Delta\zeta}$ (green) and $C_\ell^\zeta$ (red), based on a Euclid-like survey, cf. Table \ref{table:Specs}, for equal redshift correlations at $z=0.9225$. The total power is given by the solid line, the density contribution in dashed, the redshift-space distortions in dot-dashed, the other components are effectively zero. Note the pronounced BAO peaks in $C_\ell^{\Delta\zeta}$.}
    \label{fig:delta_zeta}
\end{figure}

Increasing the window width is known to decrease the total power of $C_\ell^\Delta$, since most of the power is actually from smaller scale fluctuations which are smoothed out by larger windows. In contrast, a larger window size enhances the total power of $C_\ell^\zeta$, as well as the lensing contribution to $\zeta$, since it is an integrated effect, cf. (\ref{DezNF}). Considering unequal-redshift power spectra $C_\ell(z_a,z_b)$ decreases the overall power compared to equal redshift correlations. However, the lensing contribution becomes more and more important and eventually dominates the further the redshift bins are separated. 

Studying the $\Delta$-$\zeta$ cross-spectra, see Fig. \ref{fig:delta_zeta}, shows that also here the main contribution comes from the density. Redshift-space distortions only play a role on large scales, $\ell\lesssim 100$, whereas lensing and relativistic effects are completely negligible. Furthermore, most of the multipoles are negative. What is especially interesting is that the acoustic peaks are very pronounced in this correlator. \ed{In the number counts and the $\zeta$ power spectrum, the baryon acoustic oscillation (BAO) peaks are suppressed, especially for narrow windows, since these are dominated by fluctuations on smaller scales. On the contrary, the cross-spectrum correlator has a window function with vanishing mean, which removes radial fluctuations and so we see the fluctuations at the scale $k=\ell/r(\bar z)$ without smearing it out with contributions from different wave vectors.} Therefore, the $\Delta$-$\zeta$ correlator at equal redshifts could be used as an excellent standard ruler, though it should be noted that the signal is quite small \rut{and typically has a signal-to-noise amplitude of order unity, cf. Figure \ref{fig:SNR}. Therefore, some binning in $\ell$-space will be needed for a detection. 
Since $\De(z)$ grows towards lower redshifts the mean over the redshift bin of the product $\De(z)(z-\bar z)$ is typically negative. This explains the negativity of $C_\ell^{\De\zeta}$.}

The $\Delta$-$\zeta$ correlator
actually dominates in neighboring redshift bins (see Fig.~\ref{fig:CorrCoeff} and its discussion in the next section); however, when considering unequal redshift bins, the BAO's are suppressed similarly to their suppression in the number counts and the $\zeta$ power spectrum.

\section{Fisher matrix study}\label{s:fish}
In this section, we describe the particulars of the Fisher analysis used to compare the Fisher information from the $C_\ell^\Delta$ and $C_\ell^\zeta$ angular power spectra, for two typical upcoming spectroscopic surveys. For a brief general overview of the Fisher analysis methods used, we refer the reader to appendix \ref{sec:fisherappendix}.  For our analysis, we choose fiducial cosmological parameters consistent with the Planck 2018 results \cite{planck2018}, see Table \ref{table:Cparams}, and we forecast the constraining power on these standard cosmological parameters for two different surveys. We first present the two surveys and then investigate the distribution of the Fisher information between the conventional number counts, $C_\ell^{\Delta}$, the new variable, $C_\ell^{\zeta}$, and their cross-correlation.

\begin{table}[!ht]
\centering
\begin{tabular}{c|ccccc} 
% \cline{2-3}
       Parameter & $h$& $\Omega_bh^2$ & $\Omega_ch^2$&  $n_s$&$\log(10^{10}A_s)$\\
       \hline
       Fiducial value & $0.6732$ & $0.022383$ & $0.12011$ & 0.96605 & 3.0448\\
\end{tabular}
\caption{Fiducial cosmological parameters used in the analysis. These are the best fit standard cosmological parameters
found in the Planck 2018 analysis~\cite{planck2018}. \label{table:Cparams}}
\end{table}

\subsection{The surveys}
\label{sec:surverys&specifications}
In this study we consider two spectroscopic surveys, namely a Euclid-like survey and an SKAII-like survey. While  the Euclid-like survey has many slim bins, the SKAII-like survey covers a larger range in redshift which we split into relatively wide bins. Thus, the results from two such surveys will be useful when compared, and these surveys, whose specifications (taken primarily from \cite{2020Euc} and \cite{2021SKA}) are detailed in this subsection, are taken to be representative of upcoming spectroscopic surveys. \rut{Here we are not interested in comparing the surveys directly to each other with the aim of determining which one is `better' or `worse', but rather we investigate the behaviour of the redshift-weighted galaxy number counts in response to these two different redshift binnings. We could of course choose much narrower redshift bins also for SKA.}

\begin{table}[!ht]
\centering
\begin{tabular}{ |ccc|} 
% \cline{2-3}
\hline
      Specification & Euclid-like \cite{2020Euc} & SKAII-like  \cite{2021SKA}\\ 
      
 \hline

 Linear bias $b(z)$ & $0.79 + 0.68z$ & $0.5887e^{0.8130z}$\\ 
% \hline
 Sky coverage [$deg^2$] & $15\, 000$ & $30\, 000$ \\ 
% \hline
 Redshift range & ~$0.9 < z < 1.8$~ & ~$0.1 < z <2.0$ \\ 
  Galaxy distribution & (see Eq \ref{eq:EucGalD}) & (see Ref.~\cite{2016SKA}, Table III.)\\ 
   Magnification bias & (see Eq. \ref{eq:euclidmagnbias}) & (see Eq. \ref{eq:SKAsofz})\\ 
   Number of bins & 20 & 11\\
   Redshift bin $\sigma$ & $\simeq 0.019$ & $\simeq 0.034$ --- $0.12$\\
   \hline
\end{tabular}
\caption{Comparison of general specifications of the two surveys considered\label{table:Specs}}
\end{table}
The galaxy distributions for the surveys are handled in the following way. For the Euclid-like survey, we make use of a quadratic fit to the mean number of galaxies per unit redshift per degree$^2$ from \cite{2020Euc}, for use within the given redshift interval:
\be
\frac{{\rm d}\bar N}{{\rm d}z{\rm d}\Omega} = -1494.07z^2+2598.21z+717.09\,,
\label{eq:EucGalD}
\ee
which is integrated over each bin to obtain the number of sources in each bin. For the SKAII-like survey, we perform an interpolation of the number density of galaxies given in Table III of Ref.~\cite{2016SKA}, and convert the result to a number of galaxies in each bin using the background cosmology. The numbers of galaxies in each bin are used in the calculation of the shot noise for each of the different kinds of spectra considered, as detailed in Appendix \ref{sec:shotnoise}.

\begin{figure}[!ht]
    \centering
    \includegraphics[scale=0.4]{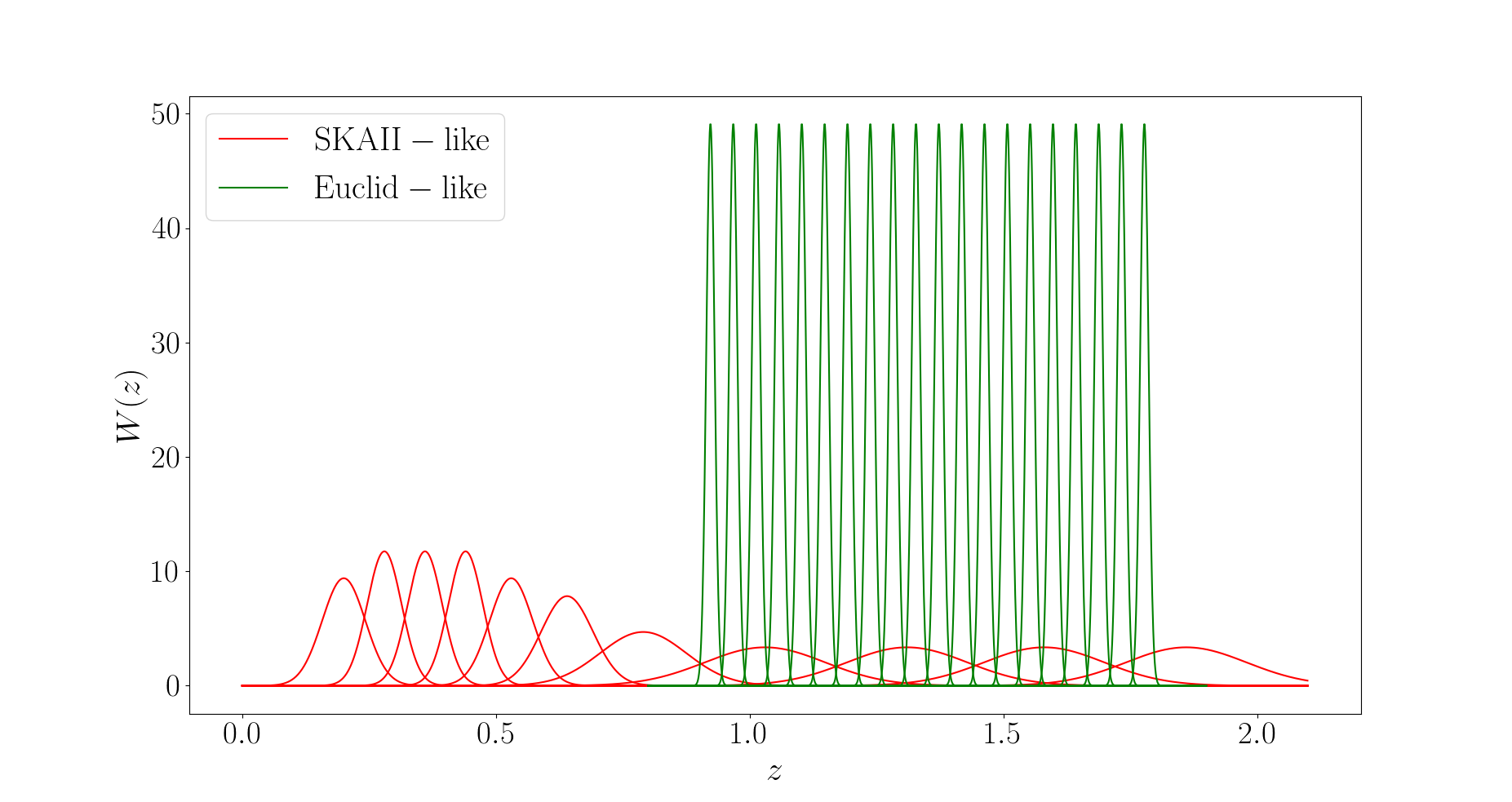}
    \caption{Redshift window functions for an 11-bin configuration of an SKAII-like survey (shown in red), and a 20-bin equal-width Euclid-like survey (shown in green).}
    \label{fig:SKAWofz}
\end{figure}

The magnification bias for the Euclid-like survey in this analysis is taken from \cite{2021Maartens}:
\be
s(z) = 0.2332 + 0.808z -0.2272z^2 + 0.01644z^3\, ,
\label{eq:euclidmagnbias}
\ee
while for the SKAII-like survey, the following fitting function from \cite{2021SKA} is used:
\be
s(z) = -0.106875 + 1.35999z - 0.620008z^2 + 0.188594z^3\, .
\label{eq:SKAsofz}
\ee
The window functions for the Euclid-like survey are defined as 20 evenly-spaced Gaussian bins. Each has a bin width (FWHM) of $\Delta z= 0.045$. I.e. we use Gaussian bins with $\sigma = \Delta z/(2\sqrt{2\ln{2}}) \simeq 0.018$.
For the SKAII-like survey, we investigate an 11-bin configuration with variable bin widths between 0.08 and 0.28. The bin widths are chosen such that all bins, except the last four, contain roughly the same number of galaxies \cite{2021SKA}, i.e., the same shot noise. The bins are also Gaussian, with sigma defined in the same way as for the Euclid-like survey. The bins are shown in Fig. \ref{fig:SKAWofz}. Due to the large bin size and relatively close spacing in redshift, there is a quite significant overlap between the bins of the SKAII-like survey. The bins are configured in such a way that the total number of galaxies, which is the number density integrated over all the bins in redshift, is the total number of galaxies expected to be observed in the survey. Galaxies with redshifts that lie in the range of overlap of two or more bins are thus assigned to only one of these bins, to conserve the total number of galaxies and prevent double-counting. This is also the reason why, despite significant overlap in redshift, the cross-correlations of different bins have vanishing shot noise, as shot noise comes from the correlation of a galaxy with itself.

%------------------------------
\rut{
The comparison of the signal amplitude (S) to the noise (N) \ed{based on the power spectra} is shown in Fig.~\ref{fig:SNR}. Taking only equal redshift data into account, the $S/N$ is given by
\begin{equation}
    \ed{\left(\frac{S}{N}\right)^2(\ell) = (2\ell+1)f_\mathrm{sky}\,C_\ell\cdot\left(\mathrm{Cov}_\ell\right)^{-1}C_\ell}\quad,
    %\frac{S}{N}(\ell) = \sqrt{(2\ell+1)f_\mathrm{sky}\,\frac{C_\ell}{C_\ell^{\mathrm{SN}}+C_\ell}}\quad,
    \label{eq:SNR_ell}
\end{equation}
where $f_\mathrm{sky}$ is the sky fraction, i.e.~the ratio between observed sky (cf.~Table \ref{table:Specs}) and the full sky. $C_\ell$ is the angular power spectrum without noise, and $\left(\mathrm{Cov}_\ell\right)^{-1}$ is the inverse diagonal covariance matrix entry for given $\ell$ and fixed redshift, given by
\begin{align}
    \left(\mathrm{Cov}_\ell\right)^{-1}&= \ed{\frac{1}{2(C_\ell^{\mathrm{SN}}+C_\ell)^2}} %\delta_{\ell\ell'}\delta_{m-m'}
    \qquad\text{for a single observable $\Delta$ or $\zeta$ ,}\\
    \left(\mathrm{Cov}_\ell\right)^{-1}&= \ed{\frac{1}{(C_\ell^{\Delta}+C_\ell^{\Delta,SN})(C_\ell^{\zeta}+C_\ell^{\zeta,SN})+\left(C_\ell^{\Delta\zeta}+C_\ell^{\Delta\zeta,SN}\right)^2}}\qquad\text{for $\Delta\zeta$ ,}
\end{align}

where $C_\ell^\mathrm{SN}$ is the shot noise contribution for the respective observable (\ref{eq:SNdelta}), (\ref{eq:SNzeta}) and (\ref{eq:SNdeltazeta}). Instead of considering the signal-to-noise ratio for a single $\ell$, one can also compute the integrated signal-to-noise ratio by summing over all $\ell$:
\begin{equation}
    \left(\frac{S}{N}\right)_\mathrm{int} = \sqrt{\sum_{\ell_\mathrm{min}}^{\ell_\mathrm{max}}\left( \frac{S}{N} \right)^2(\ell)}\quad.
\end{equation}
Figure \ref{fig:SNR} shows that, as expected, the signal-to-noise ratio is better for a configuration with wider bins (SKA-like), because of the typically higher number of galaxies per bin, as compared to slim bins. In general, because the shot noise expressions are independent of $\ell$ while the angular power spectra drop as $\ell$ increases, the signal-to-noise ratio drops with $\ell$, and eventually the noise dominates. Note that for $\Delta\zeta$, the absolute value of $\mathrm{Cov}_\ell^{\De\zeta}$ in (\ref{eq:SNR_ell}) is taken.
\begin{figure}
    \centering
    \includegraphics[width=0.55\textwidth]{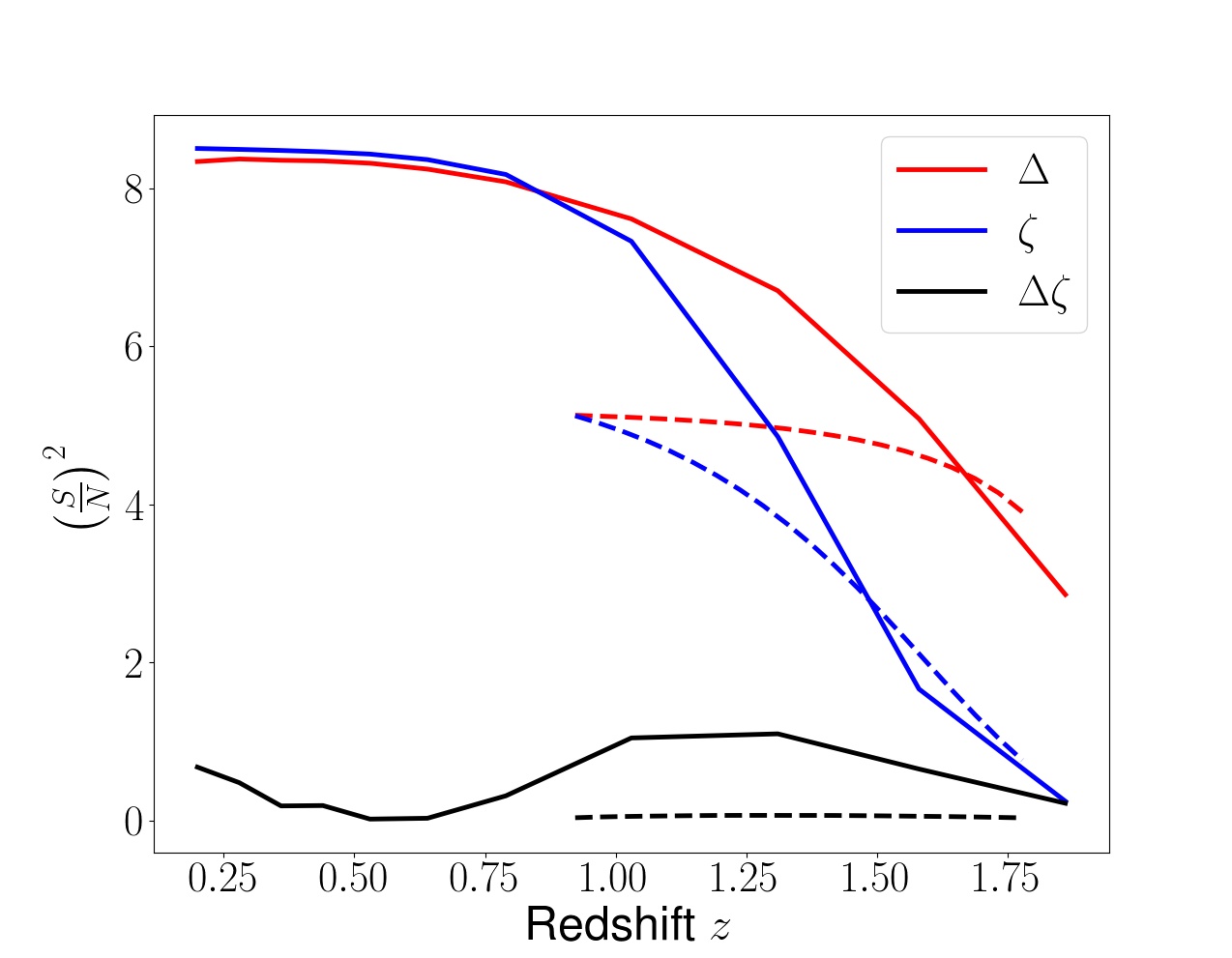}\includegraphics[width=0.55\textwidth]{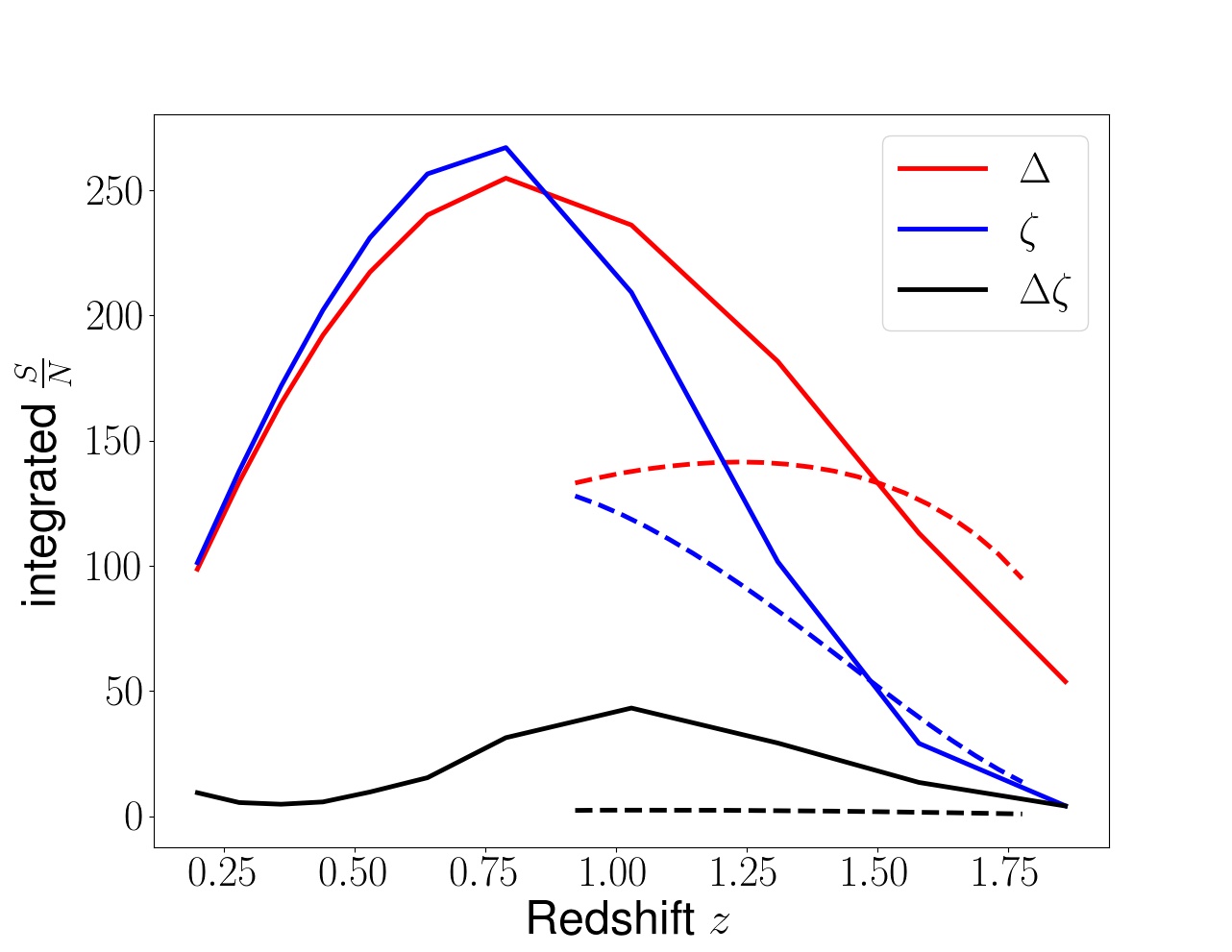}
    \caption{Left: Squared signal-to-noise ratio for fixed $\ell=100$ as function of redshift for the SKA-like configuration in solid, and for the Euclid-like configuration in dashed. Right: Integrated (summed over all $\ell$) signal-to-noise ratio for the respective surveys. In all cases considered here, the signal-to-noise ratio for $\Delta\zeta$ is \ed{a factor of several} smaller than for the individual observables.}
    \label{fig:SNR}
    %\label{fig:shotnoise}
\end{figure}}
\\ \\

In the {subsequent} Fisher analysis {in Section \ref{sec:constraints}}, we make use of the following labels for various forecasting strategies where the differences concern the included correlations, scale cut-offs, and redshift splits. Our main results will be quoted considering case "D" below, and the other strategies will be used to investigate in greater detail certain aspects of these results.%\den{maybe with a short comment for each case, why we do this?}\wil{Done! We need to make sure we discuss all these points then.}
\begin{itemize}
\item{``A" - Only equal redshift correlations of $\Delta\&\zeta$ ($\Delta-\Delta$, $\zeta-\zeta$, and cross-spectra correlations).}
\end{itemize}
By comparing with the other forecasts, forecast A allows us to examine the benefit of including cross-correlations, balanced against the disadvantage of shot noise.
\begin{itemize}
\item{``B" - All $\Delta\&\zeta$ correlations, $\ell_{\rm max} = 300$ for every bin.}
\item{``C" - All $\Delta\&\zeta$ correlations, $\ell_{\rm max} = 600$ for every bin.}
\end{itemize}
As $\ell_{\rm max}$ increases, more power from small-scale fluctuations is included in the Fisher information. By examining two different $\ell_{\rm max}$'s in forecasts B and C, we can determine how sensitive the $\zeta$ spectrum is to various scales, and whether or not an increase in the extent of the scales considered makes as dramatic an improvement to the constraints as for the $\Delta$ spectrum.
\begin{itemize}
\item{``D" - All $\Delta\&\zeta$ correlations, $\ell_{{\rm max},i} = k_{\rm max}\,r(z_i)$ for the $i^{\rm th}$ bin, where $z_i$ is the central redshift of the $i^{\rm th}$ bin and $k_{\rm max} = 0.2\,{\rm Mpc}^{-1}$. The $\ell_{\rm max}$'s for each bin and each of the surveys considered are found in Table \ref{table:lmax}. The forecast A makes use of these same cut-offs in $\ell$.

\setlength{\tabcolsep}{5pt}

\begin{table}[!ht]
\centering
\begin{tabular}{|c|ccccccccccc|} 
    \hline
    &&&&&&&&&&& \\
     {\large\bf SKAII-like}  &&&&&&&&&&& \\
       bin number &1 &2&3&4&5&6&7&8&9&10&11 \\
    $\ell_{\max}$ & 169 & 232& 292& 349& 410& 481& 569& 696& 822& 927& 1021\\ &&&&&&&&&&& \\
    {\large\bf Euclid-like} &&&&&&&&&&&\\
      bin number &1 &2&3&4&5&6&7&8&9&10&11 \\    
   $\ell_{\max}$ &     641& 664& 687& 709& 730& 751& 771& 791& 810& 829& 848\\  &&&&&&&&&&& \\
 bin number    &12&13&14&15&16&17&18&19&20 &&\\
   $\ell_{\max}$    & 866& 883& 900& 917& 933& 949& 964& 980& 995 &&\\ &&&&&&&&&&&\\%\vspace*{5pt}
   \hline
\end{tabular}
\caption{Redshift-dependent maximum $\ell$ cut-offs for the two surveys considered, based on a non-linearity scale of $k_{\rm max} = 0.2\,{\rm Mpc}^{-1}$.\label{table:lmax}}
\end{table}
\setlength{\tabcolsep}{10pt}}
\end{itemize}
While forecasts B and C extend consistently out to a fixed $\ell_{\rm max}$ for every bin, forecast D introduces a different cut-off for each bin, in an attempt to exclude any information from the non-linear regime, where the linear perturbation theory calculations are not accurate. These constraints, although sometimes containing information from perhaps fewer scales, and thus containing less information, are in fact more realistic, since our analysis is performed entirely within linear perturbation theory.

The chosen example surveys differ most considerably in their bin widths and redshift range (and thus in the number of bins). \rut{In order to investigate the effect of these characteristics, we split the SKAII-like survey into two roughly equal sets of bins, one containing those of lowest redshift and the other containing the high redshift bins most closely congruent to the 20 bins of the Euclid-like survey:}
\begin{itemize}
\item{``Low z" - SKAII-like survey's 6 lowest redshift bins with variable $\ell_{{\rm max},i}(z_i)$, as above.}
\item{``High z" - SKAII-like survey's 5 highest redshift bins (similar in range to Euclid-like survey bins) with variable $\ell_{{\rm max},i}(z_i)$, as above.}
\end{itemize}Within the SKAII-like survey this allows us to examine the effect of redshift, while keeping the galaxy density of each bin approximately constant (hence the variable bin sizes); while, comparing the High z constraints with those of Euclid, we are able to investigate the approximate effect of a factor $\lesssim 10$ change in bin width (and subsequent change in the smallest scales probed by the survey) on the information we can reliably extract from these different spectra.

%-------------------------------------------------------------
\section{Results and discussion}\label{s:res}
\rut{In this section, we first present a study of the Fisher information for both survey types in subsection \ref{sec:fisherinfo}, before discussing the constraints on the cosmological parameters obtained from a Fisher analysis based on all available information (type D) in subsection \ref{sec:constraints}. We conclude this section with a detailed study of the impact of the analysis specifications, subsection \ref{sec:detailed_study}, including the transversal cut-off $\ell_\mathrm{max}$, considering different redshift configurations, or comparing with the individual $\Delta$ and $\zeta$ constraints.

Generally speaking, by weighting the galaxy number count fluctuations $\Delta$ with redshift fluctuations in (\ref{e:exzeta}), we effectively introduce a window function with mean zero. This allows us to have a better resolution in the radial direction as compared to the original window function, which allows additional small scale information in the radial direction to be resolved. Therefore, the main advantage of studying $\zeta$ instead of $\Delta$, is that for a given survey $\zeta$ includes additional small scale information. Whether or not this additional small scale information in the radial direction eventually yields better overall constraints, will depend on how the radial resolution compares to the transversal resolution encoded in $\ell$, cf.~subsection \ref{sec:detailed_study} for more details. It is also stressed that, even if $\zeta$ provides extra small scale information, it is based on the same observational data as $\Delta$, namely galaxy number counts and redshift information. Hence, no additional measurements are required for the computation of $\zeta$. 
%main observation: scale argument
}

\subsection{Fisher information}
\label{sec:fisherinfo}
\begin{figure}[!ht]
    \centering
    \includegraphics[scale=0.4]{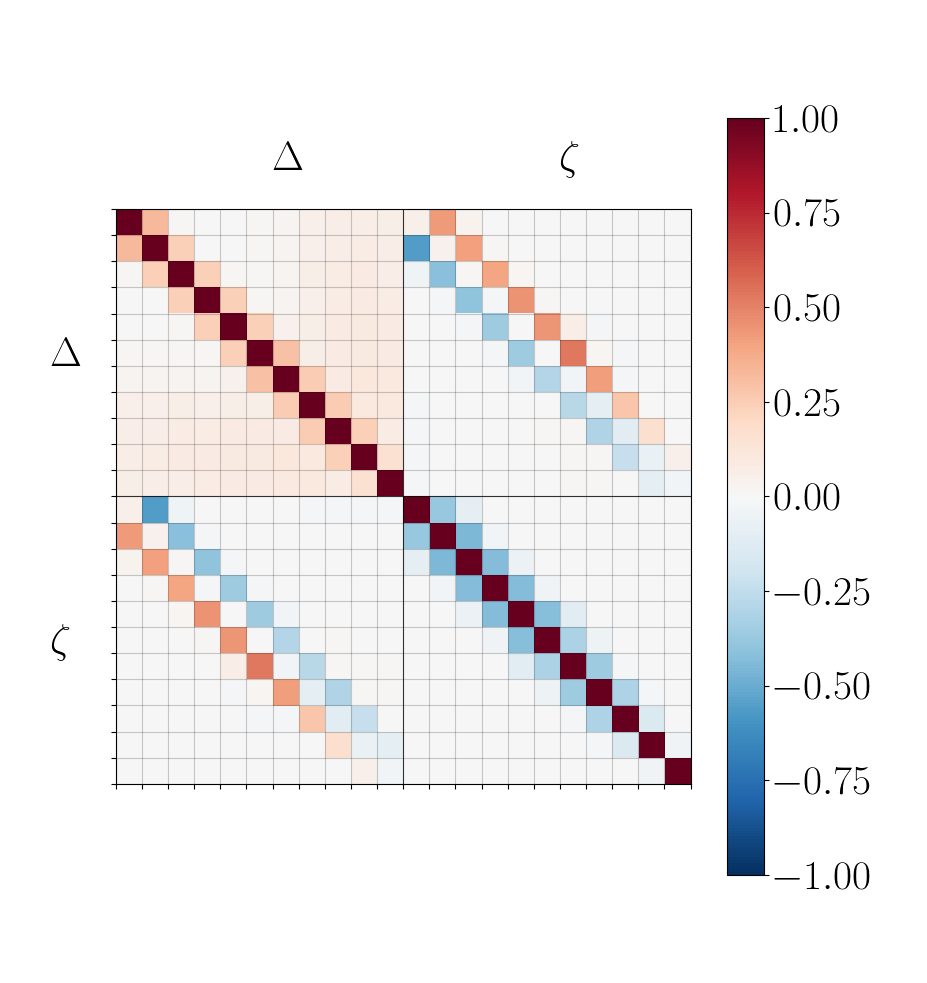}\includegraphics[scale=0.4]{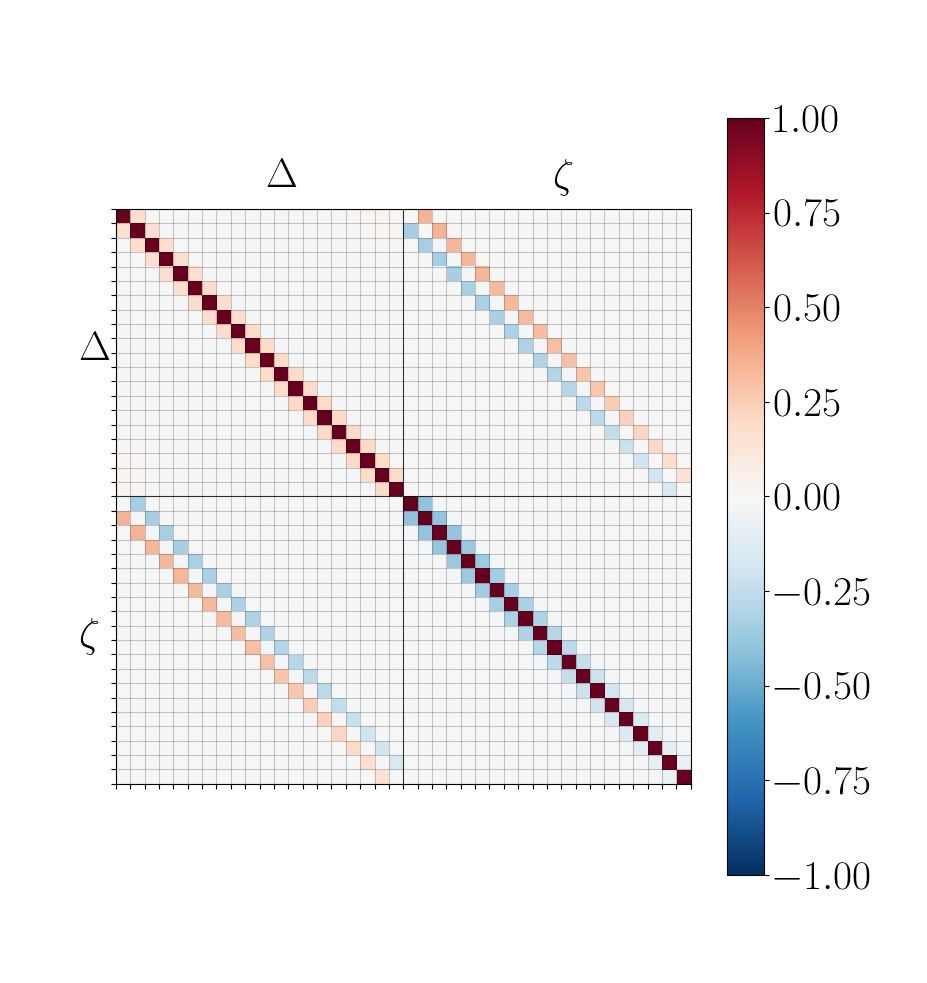}
    \caption{Correlation coefficient plots, describing \wil{the information} contained in $C_\ell^\Delta$, $C_\ell^\zeta$ and their cross-correlations at $\ell=100$, for the surveys and corresponding bins considered in the analysis. Each pixel represents a correlation coefficient between two bins. The bin redshift is increasing, and the rows and columns start in the upper left corner with the $C_\ell^{\Delta\,{\rm (obs)}}$ bins, before continuing with those of $C_\ell^{\zeta\,{\rm (obs)}}$. Values are normalised to the diagonal, and so range from $-1$ (dark blue) to $+1$ (dark red). We show our SKAII-like survey (left) and the Euclid-like survey (right).}
    \label{fig:CorrCoeff}
\end{figure}

In this section we examine the correlation matrices of each survey, to get an idea of where the most useful information is contained. Since in each survey we deal with two kinds of angular power spectra ($C_\ell^\Delta$ and $C_\ell^\zeta$), which are also calculated in the same set of redshift bins, care must be taken in interpreting the term ``auto-correlation", which will be used here exclusively in the cases where the redshift bin of a particular spectrum is correlated with itself, e.g. $C_\ell^\Delta(z_i,z_i)$. In the context of the structure of the correlation matrices presented here, this describes the entries on the main diagonal of Fig. \ref{fig:CorrCoeff}.   
%l=20
%\includegraphics[scale=0.3]{SKA-L20-CorrCoeff.png}\includegraphics[scale=0.3]{EUC-L20-CorrCoeff.png}

In Fig. \ref{fig:CorrCoeff} we show the covariance matrix  re-scaled by the auto-correlations, so that the entries correspond to the correlation coefficient of the observed spectra (including shot noise) for $\ell=100$: $\frac{C_\ell^{{\rm (obs)}\, i,j}}{\sqrt{C_\ell^{{\rm (obs)}\, i,i}C_\ell^{{\rm (obs)}\, j,j}}}$.
The off-diagonal terms in the upper left and lower right sub-matrices are single-spectrum (i.e. $\Delta\times\Delta$ and $\zeta\times\zeta$) correlations at unequal redshifts. For the $\Delta$ spectrum we observe some correlation in neighbouring bins. For low $\ell$'s,  negative correlation are also found between bins that are further apart. This comes from the negative part of the galaxy-galaxy correlation function. In the case of the SKAII-like survey, there are also some weak correlations between pairs of bins that are further apart in redshift. The same effect is not seen in the correlation coefficients of the Euclid-like survey, because the bins are very narrow which enhances the bin auto-correlations to the detriment of the cross-correlations.

Interestingly, in $\zeta$ we find negative correlations between neighbouring bins, which  decrease in strength for higher redshift bins. \rut{This decrease is most likely due to the decrease in proportion of the density and velocity components to the spectra as integrated effects like lensing become more relevant.} However, lensing is very slowly varying in redshift, so when averaged by $\zeta$ it is much less present, see also Fig. \ref{fig:contributions_euclid}.

Despite them describing the same redshift bins, the cross-correlation between $\Delta$ and $\zeta$ bins at fixed redshift (shown in the diagonals of the upper-right and lower-left sub-matrices of Fig. \ref{fig:CorrCoeff}) are very small and negative.
Moving away from the diagonals of these sub-matrices, considering the correlations between $\Delta$ and $\zeta$ at different redshifts, we see that $C_\ell^{\zeta\De}(z_i,z_j)$ is negative when $z_i< z_j$ and positive when $z_j>z_i$. This is understandable, as $\De(z_j)$ adds weight to redshifts closer to $z_j$. Hence
$\langle \De(z_j)\De(z_i)(\bar z_i-z_i)\rangle$ tends to be negative if $z_j>z_i$ and vice versa.
%they are anti-correlated. This is understandable from the definition of $C_\ell^{\Delta\zeta}(z_i,z_j)$, which may be shown to be equal to  $-C_\ell^{\Delta\zeta}(z_j,z_i)$ up to a constant multiplicative factor. \wil{I think?} \rut{why?}

The Fisher analysis takes a sum of the information from all scales, and all $\ell$'s below a certain cut-off, which may differ for bins at different redshifts. Examining the average information contained over the relevant scales, we observe an overwhelming proportion to be contained in the auto--correlations of the two angular power spectra, with only a very insignificant contribution from $\Delta\zeta$ cross correlations at mainly from unequal redshifts. This qualitative behaviour is the same for other values of $\ell$.%\wil{Plot summed over ell is available, not sure if we should add it here/mention it.}

%-------------------------------------------------------
\subsection{Constraints}
\label{sec:constraints}
\begin{figure}[!ht]
    \centering
    \includegraphics[scale=0.39]{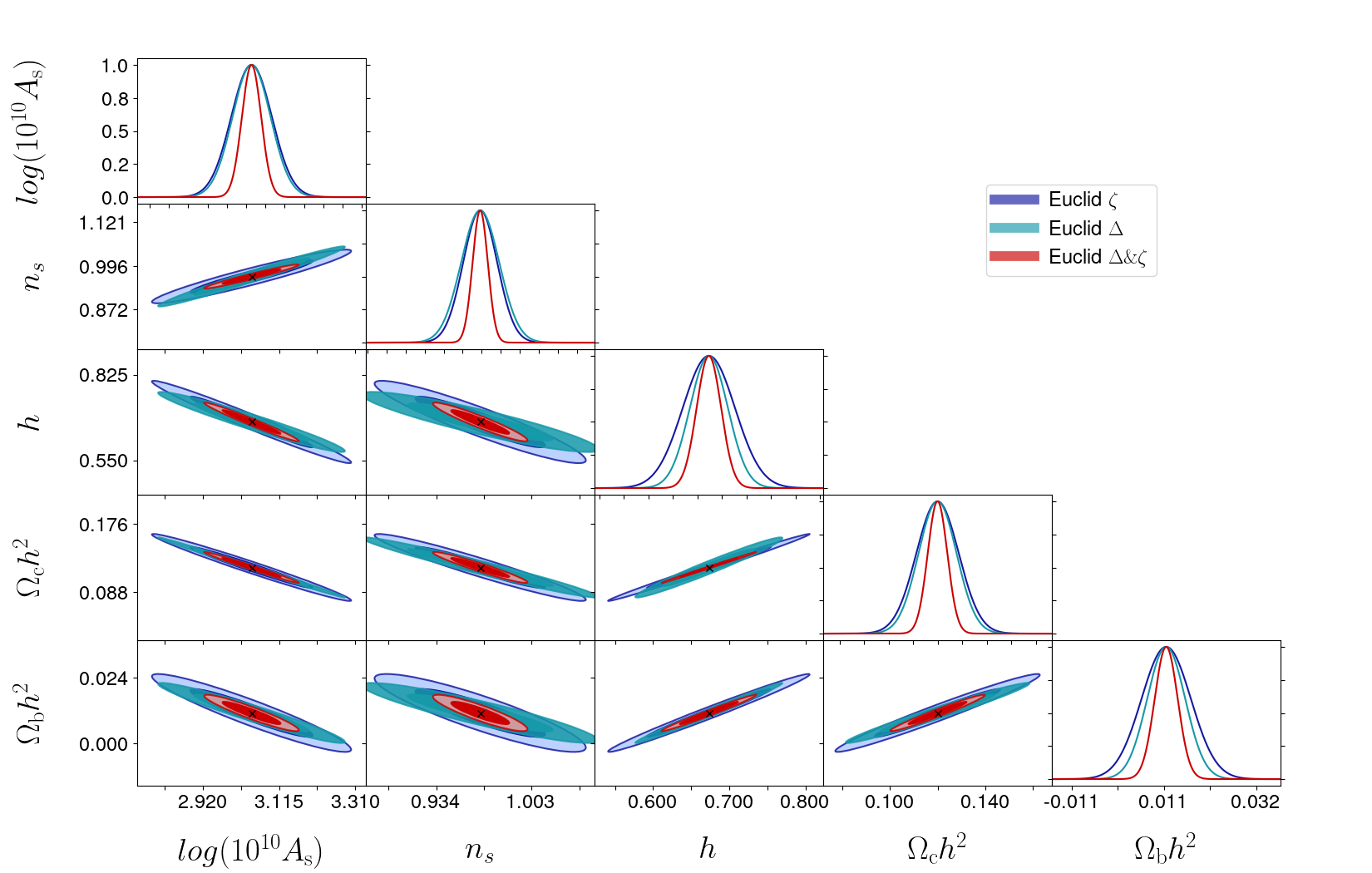}
    \caption{Constraints on the cosmological parameters based on the Euclid-like survey from Table \ref{table:Specs}. Constraints from $\zeta$ in blue, $\Delta$ in turquoise, and $\Delta\&\zeta$ in red.}
    \label{fig:triangle_euclid}
\end{figure}

\begin{figure}[!ht]
    \centering
    \includegraphics[scale=0.4]{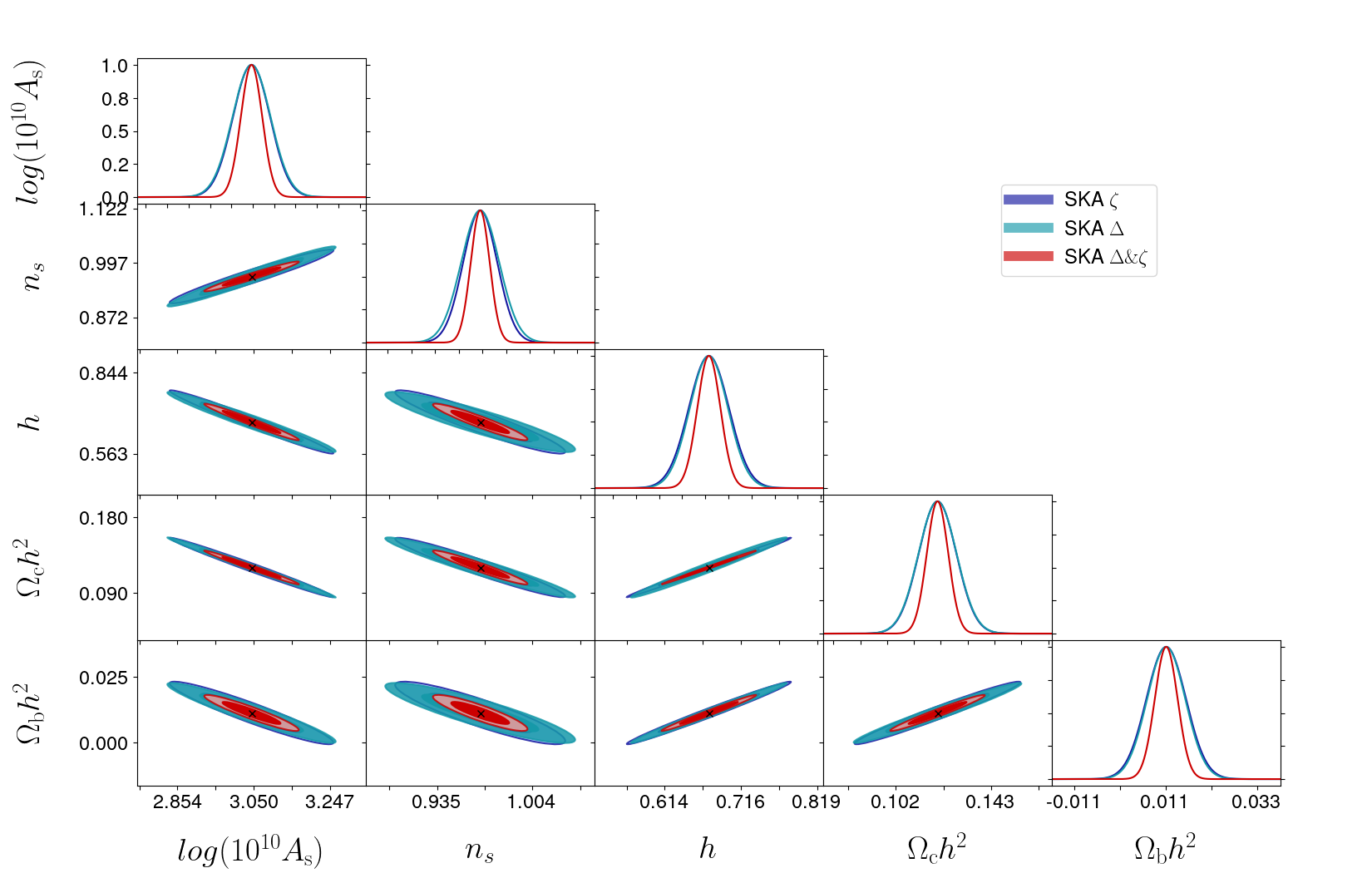}
    \caption{Constraints on the cosmological parameters based on the SKAII-like survey from Table \ref{table:Specs}. Constraints from $\zeta$ in blue, $\Delta$ in turquoise, and $\Delta\&\zeta$ in red.}
    \label{fig:triangle_SKA}
\end{figure}

The Fisher forecasts {considering all available data, cf.~type D in Section \ref{e:exzeta},} for the various spectra of the Euclid- and SKAII-like surveys are summarised in Table \ref{table:finalconstr} and visualised in Figures \ref{fig:triangle_euclid} and \ref{fig:triangle_SKA}. 
We first note that in both cases, the constraints based on $\zeta$ alone are more often than not slightly worse than those from $\Delta$ alone, on average by about 5\% for the SKAII-like survey and by about 20\% for the Euclid-like survey. However, for the Euclid-like survey, $n_s$ is better determined by the $\zeta$ analysis than by $\De$. The SKAII-like survey  leads to almost identical constraints from the $\De$ analysis alone and from the $\zeta$ analysis alone.  As expected, the best constraints come from the combined $\zeta\&\Delta$ analysis, demonstrating 
%This result contrasts strongly with the previous findings in \cite{Hernandez-Monteagudo:2020xnl}, but it is important to note that there are several differences in the analysis parameters, most notably the maximum $\ell$ considered, the consequences of which will be investigated in the next subsection. \den{I formulated a small paragraph on the comparison with Louis at the end of this section.}
%\den{switched the order}
that the new observable $\zeta$, which may be directly determined from the same set of galaxy survey redshifts and angular positions as the $C_\ell^\Delta$ spectrum, contains useful additional information.

The improvement garnered from the combined spectra, compared to constraints from the $C_\ell^\Delta$ spectrum alone, is around $45\%$, with the Euclid-like survey showing a slightly more substantial improvement than the SKAII-like survey. As can be seen in Fig.~\ref{fig:triangle_euclid}, for most parameters there is a misalignment of the ellipses for $\Delta$ and $\zeta$, meaning that considering $\zeta$ in addition to $\Delta$ helps to break  degeneracies of the $\Delta$ constraints. In contrast, the ellipses for SKA with wider bins, see  Fig.~\ref{fig:triangle_SKA}, are practically aligned and so the improvement on the overall constraints comes mainly from combining the individual statistics.
%For the set of parameters chosen for these type D forecasts, we can see that this improvement must derive mainly from the strength of the individual $C_\ell^\Delta$ and $C_\ell^\zeta$ constraints, rather than a breaking of degeneracies, since the error ellipses for these individual spectra in figures \ref{fig:triangle_euclid} and \ref{fig:triangle_SKA} are almost completely aligned.

When comparing the final $\Delta\&\zeta$ constraints from each of the surveys, we notice that the Euclid-like survey's forecasts are very slightly better. However, considering the level of accuracy in the Fisher analysis, we do not believe that this small difference is significant. Furthermore, the constraints from the individual $C_\ell^\Delta$ and $C_\ell^\zeta$ spectra are actually slightly worse for the SKAII-like survey.
This is not so surprising as the two surveys are significantly different in their binning strategies, and consequently exhibit different behaviours in their forecast constraints. These differences, including the dependence of constraints on the bin width, redshift and relevant scales, will be investigated in further detail in the subsection to follow.

It is also interesting to look at the degradation of the final result from the total combined spectra when only the combination of the individual spectra $(\Delta+\zeta)$, but not the cross $\Delta\zeta$ correlations, are used. For the Euclid-like survey we see a degradation of around $4.5\%$ on average, while for SKAII-like survey, there is a larger loss in constraining power - about $14\%$. This seems to indicate that the large, significantly overlapping bins of the SKAII-like survey lead to a larger contribution in not only off-diagonal (i.e. different redshift) bins of each of the individual spectra, but also in the cross-correlations of $\Delta\zeta$, which is largest in unequal redshift correlations.

The above results contrast strongly with the previous findings in \cite{Legrand:2020sek}: the constraints for the Euclid-like survey presented here are somewhat better for the $C_\ell^\Delta$ spectrum than the $C_\ell^\zeta$ spectrum, while in \cite{Legrand:2020sek} the reverse is true, despite the survey specifications being almost identical.\footnote{\rut{However, one major difference in the forecasting approaches of this paper and \cite{Legrand:2020sek} is that here we do not marginalise over the bias parameters.}} But it is important to note that there are several differences in the analysis. One of the major differences in our analysis is that instead of a fixed cut-off at $\ell_{\rm max} = 300$, we include smaller scales in our bins at higher redshift, according to $\ell_{\rm max}(z) = 0.2\,{\rm Mpc^{-1}}r(z)$, which translates into $\ell_{\rm max}=169-1021$ for SKA and $641-995$ for Euclid, cf. survey type D in Table~\ref{table:lmax}. Indeed, while our results, which include more small-scale information for each bin in a redshift-dependent manner, display tighter constraints, Table~\ref{table:finalconstr} shows that the constraining power of $C_\ell^\zeta$ in this case is a bit weaker than that of $C_\ell^\Delta$. It will therefore be instructive to study the $\ell_{\rm max}$-dependence of the constraints in greater detail, and we do this in the next subsection. We also arrive at slightly different shot noise expressions (\ref{eq:SNzeta}) \& (\ref{eq:SNdeltazeta}) than the ones given in Ref.~\cite{Legrand:2020sek}, see Appendix~\ref{sec:shotnoise} for details.

\begin{table}[!ht]
\centering
\begin{tabular}{c|c|ccccc} 
% \cline{2-3}
       Survey & Forecast type & $\sigma[h]$& $\sigma[\Omega_bh^2]$ & $\sigma[\Omega_ch^2]$&  $\sigma[n_s]$&$\sigma[\log(10^{10}A_s)]$\\
       \hline
       \multirow{4}{*}{Euclid-like} & $\zeta$ &  0.05327& 0.005719&0.01733&0.03094&0.1026 \\
       & $\Delta$ &  0.03889& 0.004413&0.01556&0.03465&0.09614 \\
       & $\Delta+\zeta$ & 0.02613 & 0.002791 & 0.008372 & 0.01449& 0.05078 \\
         & \CC $\Delta$\&$\zeta$ & \CC 0.0250& \CC 0.002670& \CC 0.00801& \CC 0.01384& \CC 0.0487 \\
       \hline 
       \multirow{4}{*}{SKAII-like} & $\zeta$ & 0.04441 & 0.004797 &0.01433 &0.02492 &0.08522 \\
       & $\Delta$ & 0.04218 & 0.004607 &0.01451 &0.02782& 0.08762\\
       & $\Delta+\zeta$ & 0.02926 & 0.003155& 0.009353& 0.0158 & 0.05617\\
       &\CC $\Delta\&\zeta$ & \CC 0.0256 & \CC 0.002761& \CC 0.00817 &\CC 0.01385 &\CC 0.0492\\
       \cline{2-7}
\end{tabular}
\caption{Final (type D) marginalised constraints on the background cosmological parameters, forecast based on $C_\ell^\Delta$, $C_\ell^\zeta$ and $C_\ell^{\Delta\&\zeta}$ angular power spectra for Euclid-like and SKAII-like surveys. Note that `$\Delta+\zeta$' here refers to the forecasts from the matrix of diagonal submatrices, i.e. the total $\Delta\&\zeta$ without $\Delta\zeta$ cross-correlations.}\label{table:finalconstr}
\end{table}

%--------------------------------------------
\subsection{Dependence of the constraints on the analysis specifications}
\label{sec:detailed_study}

The final results presented in the previous subsection represent the most realistic combination of analysis parameters of the various types of forecasts considered. We use as much information as possible (including all the cross-correlations between different spectral bins and bins at different redshifts) and, for each bin, we only include information from the corresponding scales where we deem the linear perturbation theory used in this analysis to hold.\footnote{Strictly speaking, we should actually consider the additional effect of narrow window bins bringing in non-linear radial scales which propagate throughout all angular scales in $\ell$, as described in \cite{Matthewson:2021rmb}} The constraints of the full complement of forecasts are presented in Table \ref{table:finalconstr2}, for the combination $\Delta\&\zeta$ of $C_\ell^\Delta$ and $C_\ell^\zeta$. As mentioned above, the results from this realistic analysis show that the angular power spectrum of redshift fluctuations, $C_\ell^\zeta$, is a useful observable in conjunction with the more widely-known and used $C_\ell^\Delta$. Before commenting on the individual survey types considered in Table \ref{table:finalconstr2}, we give an argument under which circumstances $\zeta$ might yield better constraints than $\Delta$.

An important point is that, depending on the competing resolutions in radial and transversal directions, additional small-scale information may or may not be gained by $\zeta$. More specifically, the transversal \rut{resolution is set by the minimal angular extent $\theta_{\min}\simeq \pi/\ell_\mathrm{max}$, which can also be expressed as a comoving transversal scale $D_\perp= r(z)\frac{\pi}{\ell_\mathrm{max}}$}, and is the same for both $\Delta$ and $\zeta$. The radial resolution on the other hand, is related to the scale, $D_\parallel$, probed in the radial direction connected to redshift. For $\Delta$, this radial scale is determined by the width of the redshift bin $\Delta z/H(z)$, whereas $\zeta$ probes smaller radial scales. The weighting by the redshift fluctuations in the integral in (\ref{e:exzeta}) ensures that, while the integral $\int (\bar z -z_i)\om(z,z_i)dz =0$, the $\zeta$ variable is measuring radial fluctuations on scales smaller than $\Delta z$, typically $\De z/2$, which translates into $D^\zeta_\parallel=\frac{\Delta z}{2 H(z)}$. \rut{In order for $\zeta$ to add additional small scale information, the condition $D_\parallel^\zeta<D_\perp$ has to be satisfied. In Table \ref{table:lrad}, we compare the transversal resolution $D_\perp\simeq \pi/0.2 {\rm Mpc}=15.71$Mpc, which is the same for both surveys in our case, versus the radial resolution $D^\zeta_\parallel$ of $\zeta$. For both survey configurations it turns out that the transversal resolution is always better than the \ed{radial} resolution: $D_\perp<D^\zeta_\parallel$. In other words, the $\zeta$ observable cannot resolve smaller scales beyond the smallest scales resolved in transversal direction. Therefore, we do not expect that $\zeta$ gives better constraints on the cosmological parameters than $\Delta$, which is in qualitative agreement with Figures \ref{fig:triangle_euclid} \& \ref{fig:triangle_SKA}. } \\

\setlength{\tabcolsep}{5pt}

\begin{table}[!ht]
\centering
\begin{tabular}{|c|ccccccccccc|} 
    \hline
    &&&&&&&&&&& \\
     {\large\bf SKAII-like}  &&&&&&&&&&& \\
       bin number &1 &2&3&4&5&6&7&8&9&10&11 \\
    $D_\parallel^{\zeta}\, [{\rm Mpc}]$ & 85.29& 65.22& 62.25& 59.53& 70.24& 78.81& 119.85& 145.36& 123.93& 107.29&93.33\\
    &&&&&&&&&&& \\

    %$\De_\perp [{\rm Mpc}]$ &    \multicolumn{11}{c|}{15.71}\\
    %&&&&&&&&&&& \\
    {\large\bf Euclid-like} &&&&&&&&&&&\\

 bin number    &1&2&3&4&5&6&7&8&9 &10&11\\
  $D_\parallel^{\zeta}\, [{\rm Mpc}]$  & 24.90&24.24& 23.60&22.99& 22.40&21.82& 21.27&20.74& 20.22& 19.73& 19.25\\%\vspace*{2pt}
bin number    &12&13&14&15&16&17&18&19&20 &&\\
  $D_\parallel^{\zeta}\, [{\rm Mpc}]$  &  18.78& 18.34&17.91& 17.49& 17.09& 16.70& 16.33& 15.96& 15.61&&\\  &&&&&&&&&&&\\%\vspace*{5pt}      
      
%    $\De_\perp [{\rm Mpc}]$  &  \multicolumn{11}{c|}{15.71}  \\ 
%    &&&&&&&&&&& \\
 \hline
\end{tabular}
\caption{\rut{Scale of the radial resolution $D_\parallel^{\zeta}$ associated with the $\zeta$ observable for each bin in the respective survey. Notice that the radial resolution $D_\parallel^{\zeta}$ is always worse than the transversal resolution $D_\perp=15.71$Mpc, apart from the last bin in the Euclid-like survey. Hence, for these configurations, $\zeta$ does not contain additional new small scale information.}\label{table:lrad}}

\end{table}
\setlength{\tabcolsep}{5pt}

\rut{\subsubsection{Equal redshift data forecasts}}
The first set of additional forecasts, type A, are made with the same redshift-dependent $\ell_{\rm max}$ as type D, but including only the correlations of equal redshift bins for $C_\ell^\Delta$, $C_\ell^\zeta$ and $C_\ell^{\Delta\zeta}$. Neglecting correlations coming from different redshifts decreases the total information available; hence, the constraints are expected to degrade. This is indeed what is observed in Table \ref{table:finalconstr2}, which can also be understood by looking again at the correlation matrices in Figure \ref{fig:CorrCoeff}. Equal redshift correlations are located along the diagonal of each sub matrix; thus, for type A forecasts, all other correlations, some of which are clearly non-zero, are neglected. Overall, this degradation is much more pronounced for the SKAII-like survey, and the constraints are on average around $30\%$ weaker than when the full information is used. For the Euclid-like survey the constraints degrade by only about $10\%$. This is easily understood, since between the wide redshift bins used in the SKAII-like analysis there is considerable overlap, especially between neighboring bins.  This is much less significant for the slim bins of the Euclid-like analysis
(see Fig.~\ref{fig:SKAWofz}).
It is also interesting to compare the constraints from correlations at equal redshift (type A) to those from the full fisher information (type D), in the cases of the individual $C_\ell^\Delta$ and $C_\ell^\zeta$ spectra of each survey. The $\Delta\zeta$ correlations at equal redshifts are very small, and so most of the information is coming from the off-diagonal entries.
For the Euclid-like survey, the degradation in constraints from each of the individual spectra is fairly similar, while for the SKAII-like survey the degradation is much more apparent in the case of the individual $C_\ell^\zeta$ spectrum. Seemingly, when there is a significant overlap in the redshift bins, the information of the $C_\ell^\zeta$ spectrum is shifted more into the correlations between different redshifts. %\wil{I'm not sure if we should try to understand this, but perhaps with larger bins, the RSD is less significant in autocorrelations of $\Delta$ (and so even in $\zeta$ the enhancement is not so large) and so $\zeta$ performs a useful role in the unequal redshift correlations, by increasing the velocity information?}

%However, comparison with the existing literature, in particular \cite{Hernandez-Monteagudo:2020xnl}, prompted further investigation of the characteristics of the constraints and their dependence on various parameters of the analysis. In particular, we noticed that the constraints for the Euclid-like survey presented here are better for the $C_\ell^\Delta$ spectrum than the $C_\ell^\zeta$ spectrum, while in \cite{Hernandez-Monteagudo:2020xnl} the reverse is true, despite the survey specifications being almost identical. One of the major differences in our analysis is that instead of a fixed cut-off at $\ell_{\rm max} = 300$, we include smaller scales in our bins, according to $\ell_{\rm max}(z) = 0.2\,{\rm Mpc^{-1}}\chi(z)$. The effect of this redshift-dependent cut-off seems to suppress the ability of $C_\ell^\zeta$ to constrain the cosmological parameters to a better degree than $C_\ell^\Delta$. Indeed, while our results that include more small-scale information for each bin in a redshift-dependent manner display tighter constraints \wil{(Is this really significantly the case? Our results for D-type are actually quite similar to Louis')}, Table \ref{table:finalconstr} shows that the constraining power of $C_\ell^\zeta$ in this case is weaker than that of $C_\ell^\Delta$. 
\vspace{0.2cm}

\rut{\subsubsection{Effect of the transversal cut-off scale $\ell_\mathrm{max}$}}
In the forecasts of type B, we investigate the effect of a constant $\ell = 300$ \rut{($D_\perp \sim \frac{r(z)}{100}$, see Table \ref{table:300Dperp})} cut-off across all bins. The Euclid-like survey's constraints from this kind of forecast are comparable to those from \cite{Hernandez-Monteagudo:2020xnl} and, more importantly, the constraints from the individual $C_\ell^\zeta$ spectrum are stronger than those from the individual $C_\ell^\Delta$ spectrum. This is not the case for the SKAII-like survey and can be understood by noting the following (compare Tables \ref{table:lrad} and \ref{table:300Dperp}): while for the Euclid-like survey we typically have \rut{$D_{\rm \parallel}<D_{\perp}(\ell=300)$, so that improving on the radial information using $\zeta$ truly includes smaller modes in the analysis, this is not the case for the SKAII-like survey, where most bins have $D_\parallel>D_\perp(\ell=300)$. }
The effect is even more pronounced in the Euclid-like analysis of Ref.~\cite{Hernandez-Monteagudo:2020xnl} since it considers bins which are about a factor 2 more narrow (they have $\si_z=0.01$).

\setlength{\tabcolsep}{5pt}

\begin{table}[!ht]
\rut{
\centering
\begin{tabular}{|c|ccccccccccc|} 
    \hline
    &&&&&&&&&&& \\
     {\large\bf SKAII-like}  &&&&&&&&&&& \\
       bin number &1 &2&3&4&5&6&7&8&9&10&11 \\
    $D_\perp\, [{\rm Mpc}]$ & 8.87& 12.17& 15.31& 18.31& 21.51& 25.20& 29.84& 36.44& 43.07& 48.55& 53.48\\
    &&&&&&&&&&& \\

    %$\De_\perp [{\rm Mpc}]$ &    \multicolumn{11}{c|}{15.71}\\
    %&&&&&&&&&&& \\
    {\large\bf Euclid-like} &&&&&&&&&&&\\

 bin number    &1&2&3&4&5&6&7&8&9 &10&11\\
  $D_\perp\, [{\rm Mpc}]$  & 33.60& 34.81& 35.99& 37.14& 38.26& 39.35& 40.41& 41.45& 42.46& 43.45& 44.41 \\%\vspace*{2pt}
bin number    &12&13&14&15&16&17&18&19&20 &&\\
  $D_\perp\, [{\rm Mpc}]$  &  45.34& 46.26& 47.15&48.03& 48.88& 49.71& 50.53&51.32& 52.10&&\\  &&&&&&&&&&&\\%\vspace*{5pt}      
      %    $\De_\perp [{\rm Mpc}]$  &  \multicolumn{11}{c|}{15.71}  \\ 
%    &&&&&&&&&&& \\
 \hline
\end{tabular}
\caption{\rut{Scale of the transversal resolution $D_\perp$ associated with the $\Delta$ and $\zeta$ observables for each bin in the respective survey, when the maximum multipole considered is $\ell=300$.}\label{table:300Dperp}}
}
\end{table}
\setlength{\tabcolsep}{5pt}

In order to further investigate the dependence of the constraints on the scales considered, we also prepared forecasts of type C, which extend the maximum angular multipole to $\ell_{\rm max} = 600$. We acknowledge that, on the grounds that the analysis is wholly dependent on linear perturbation theory, these constraints will not be realistic, extending in some bins into scales which are firmly in the non-linear regime. However, since we make use of a redshift dependent $\ell_{\rm max}$ in the final results, it is a useful investigation to determine the overall scale-dependence of the constraints from the angular power spectra. In both surveys, the additional scale information improves constraints by around $45\%$, and in the case of the Euclid-like survey these added scales seem to provide more additional information to the  $C_\ell^\Delta$ spectrum. That is to say, once we include all scales up to $\ell = 600$ consistently for all bins, the constraints from the $C_\ell^\Delta$ spectrum alone are not much weaker than $C_\ell^\zeta$ spectrum constraints in the case of the Euclid-like survey. For the SKAII-like survey the individual $C_\ell^\Delta$ spectrum once again has more constraining power than the $C_\ell^\zeta$ spectrum. This is consistent with the argument that the scales on which $C_\ell^\zeta$ provides more information compared to $C_\ell^\Delta$ have already been surpassed at $\ell=300$, \rut{(i.e. $D_\perp$ now considers scales even smaller than before and so we still have $D_\parallel > D_\perp(\ell=600)$). }

The relevant proportion of the contributions of velocity (RSD and RSD-Density) components to the total spectrum (at various scales) is also important. There is a play-off between two effects that this has on the spectra: (1) The wider the bins in redshift, the lower the amplitude and proportion of RSD information in the $C_\ell^\Delta$ spectrum, and so $C_\ell^\zeta$ has a higher chance of bringing relevant information from RSD. However, (2) the enhancement provided by $C_\ell^\zeta$ to the proportion of the velocity information is limited by the radial scale probed by the bins. Thus thinner bins might be needed to have a proportion of RSD in the $C_\ell^\Delta$ spectrum that is sufficiently large initially so that $C_\ell^\zeta$ is able to provide useful additional information. 

In the case of the Euclid-like survey's type C forecasts, even though the individual spectra have very similar constraints, the improvement over $C_\ell^\Delta$ of the combination of the two spectra is still significant, about $55\%$ on average. This indicates that the information that the $C_\ell^\zeta$ spectrum provides is useful in breaking degeneracies in the measurements of the cosmological parameters from $C_\ell^\Delta$. If we examine the corresponding triangle plots for these constraints we observe that the error ellipses are not aligned, similar to the type D forecasts, where we vary $\ell_{\rm max}$, see Fig.~\ref{fig:triangle_euclid}. 
%\wil{?In the case of Euclid, we always have at least up to $\ell = 640$, so we only ever include EXTRA scales. Although the combined spectra constraints (D) end up being tighter ($\sim10\%$) than type C, we lose the "misalignment". This indicates that zeta does not have as much novel information on these extra scales, compared to Delta. (Do we know why, and do we mention the radial/transverse argument here (again)?}\\

The constraints from type C forecasts of the SKAII-like survey behave  differently. As already noted for $\ell_{\rm max} = 300$ (type B), the individual spectra constraints are stronger for $C_\ell^\Delta$ than $C_\ell^\zeta$, and the error ellipses are almost completely aligned. Correspondingly, the improvement to the $C_\ell^\Delta$ constraints from combination with $C_\ell^\zeta$ is only $\lesssim 30\%$. As the bins are wider, the transverse scales already contain  the additional radial information, resulting from $\zeta$. But nevertheless, since the $\De$ and $\zeta$ window functions are nearly orthogonal, the information from $\zeta$ can still be added to the one from $\De$ to significantly improve the constraints.

\rut{\subsubsection{Effect of low \& high redshift data and the bin width}}
Since the SKAII-like survey differs from Euclid in two main respects, namely the redshift range and the bin widths, we also investigated the constraints that can be obtained if the bins of the SKAII-like survey are separated into low- and high-redshift subsets. The 5 bins with highest redshift correspond approximately to the redshift range of  the Euclid-like survey, see Fig. \ref{fig:SKAWofz}. They form our High z sample. The remaining bins form the Low z sample. In all bins we choose $\ell_{\max}$ according to the linearity condition, as for analysis D. If we compare these constraints, we observe that the Low z forecasts are about $15\%$ better than the constraints from the High z sample, see Table~\ref{table:finalconstr2}. 
On the one hand, the Low z sample has one more bin than the high z, but on the other hand it has significantly fewer independent $\ell$-modes. The total number of $\ell$-modes in the Low z sample is 1447, while the High z sample contains 4512  $\ell$-modes. The main reason for why the Low z sample nevertheless gives better constraints is the width and significant overlap of the  High z bins in addition to the significantly higher shot noise in these bins.
 Compared to the forecast using all the bins,  neglecting close to half the bins at low or high redshift (and their cross-correlations) degrades the constraints by between $\sim25\%$ and $35\%$. In the case of the High z sample, the individual $C_\ell^\Delta$ constraints are consistently better than the $C_\ell^\zeta$ constraints, while the reverse is true in the Low z forecast. The improvements to the individual spectra resulting from the use of the full set of 11 bins (which, to remind the reader, has no overwhelming preference for either of the individual spectra in the parameter constraints) are thus greater for $C_\ell^\zeta$ in the High z case and greater for $C_\ell^\Delta$ in the Low z case.\\

Since they are similar in redshift range, it is also interesting to compare the High z sample of the SKAII-like survey to the type D forecasts of the Euclid-like survey. \rut{In this way (up to the difference in the linear biases of the surveys) we can gain some insight into the effects which the number of bins has on the constraints.} The first point to note is that the constraints from the High z SKAII-like survey are about $25\%$ worse than the ones obtained with many slim bins in the Euclid-like survey, see Table~\ref{table:finalconstr2}. Based on our previous arguments  this makes sense, since there is greater scope for additional information from $C_\ell^\zeta$. Interestingly, if we examine the improvements from decreased bin size and increased bin number in the constraints from individual spectra, those from the $C_\ell^\zeta$ spectrum alone are marginally stronger, about $36\%$, than the $29\%$ improvement in the individual $C_\ell^\Delta$ spectrum. This agrees with the assertion that $C_\ell^\zeta$ is more responsible for the providing the information that results in the improvement seen in the $\Delta\&\zeta$ constraints. It also makes sense that the constraints from both individual spectra show some improvement, since the main difference between Euclid-like D and SKAII-like High z is the size of the bins (and thus the radial scales probed), from which both spectra $C_\ell^\Delta$ and $C_\ell^\zeta$ profit. We also perform B and C type forecasts on these reduced-bin samples and observe that, for the High z constraints, the error ellipses of the individual spectra are very similar while the Low z constraints for the $\De$-spectra are much better than those from the $\zeta$-spectra. Here again, $\ell_{\max}\gg \ell_{\rm rad}$ so that the additional radial information from $\zeta$ is not very significant. This is true for the High and Low z surveys; however, the Low z $\zeta$-spectra have a much lower amplitude which leads to the worsening of the constraints.

\begin{table}[!ht]
\centering
\begin{tabular}{c|c|ccccc} 
% \cline{2-3}
       Survey & Forecast type & $\sigma[h]$& $\sigma[\Omega_bh^2]$ & $\sigma[\Omega_ch^2]$&  $\sigma[n_s]$&$\sigma[\log(10^{10}A_s)]$\\
       \hline
        & A &  0.02746& 0.002947&0.008782&0.01512&0.05311 \\
       Euclid-like constraints
       & B &  0.04934& 0.005075&0.01596&0.02958&0.09877 \\
   ($\Delta\&\zeta$)     & C &  0.02789& 0.002951&0.008959&0.01593&0.05466 \\
        & \CC D &  \CC 0.025& \CC 0.00267&\CC 0.008011&\CC 0.01384&\CC 0.04867 \\ 
        \hline
    & A & 0.03368 & 0.003616 &0.01076 &0.01808& 0.06463\\
  SKAII-like constraints     & B & 0.02942 & 0.00321 &0.009377 &0.02942& 0.05587\\
  ($\Delta\&\zeta$)     & C & 0.01645 & 0.001934 &0.005079 &0.006746 & 0.02941\\
       & \CC  D & \CC 0.02558 & \CC 0.002761 &\CC 0.008175 &\CC 0.01385& \CC 0.0492\\
       \cline{2-7}
       & Low z & 0.03387 & 0.00367 &0.01086 &0.01837& 0.06553\\
       & High z & 0.03963 & 0.004248 &0.01266 &0.02165& 0.07629\\
\end{tabular}
\caption{Final marginalised constraints on the background cosmological parameters, forecast for the combination of $C_\ell^\Delta$,  $C_\ell^\zeta$ and $C_\ell^{\Delta\zeta}$ angular power spectra ($\Delta\&\zeta$) for each of the surveys considered, in several different cases described in the text and briefly recapitulated here - A: Only equal redshift correlations, B: $\ell_{{\rm max},i} = 300$, C: $\ell_{{\rm max},i}=600$, D: $\ell_{{\rm max},i} = 0.2\,r(z_i)/$Mpc, Low z: SKAII-like survey's 6 lowest redshift bins, High z: SKAII-like survey's 5 highest redshift bins. \label{table:finalconstr2}}
\end{table}
${ }$\\
${ }$\\

%----------------------------------------------------------
\section{Conclusions}\label{s:con}
In this paper we have introduced a new observable called `redshift-weighted galaxy number counts' which can be constructed from number count data. A similar quantity was called `redshift fluctuation' in previous literature~\cite{Hernandez-Monteagudo:2019epd,Hernandez-Monteagudo:2020xnl,Legrand:2020sek} but the previous definition was not on a solid footing within relativistic cosmological perturbation theory. Here we defined a variable which is gauge-invariant and which, in a quasi-Newtonian situation - as is the case in a $\La$CDM cosmology, reduces nearly to the one used in previous literature.
Additional terms from lensing and large-scale relativistic contributions are even smaller than for galaxy number counts since the contribution from terms which are constant in redshift vanishes and both lensing and large-scale relativistic contributions vary very slowly with redshift.

We have also studied the gain this variable can provide for cosmological parameter estimation from galaxy number counts. Since precise redshift measurements are needed, we concentrate on spectroscopic surveys. More precisely, we consider a Euclid-like survey and an SKAII-like survey. We find that the additional information from the new variable which we call $\zeta$ leads in all cases to a significant improvement of the constraints, from 30\% to more than 50\%, depending on the analysis. In particular, when slim redshift bins are considered, the parameter constraints from $\zeta$ are typically better than, or at least on the same level as, the ones from the number count spectra. 
They are also sure to be better if low angular resolution, low $\ell_{\max}$, is considered. If the redshift bin is very wide, the $\zeta$ variable contributes not new, but nevertheless nearly independent, information. In this case it improves parameter constraints typically by about 30\%. This is especially interesting since this new observable can be obtained `for free' from a number count survey. No additional data is needed, since it is just a different analysis of the number count data with a special `window function' which has a vanishing redshift integral.

In this paper we have considered a very simple analysis with fixed bias $b(z)$ and varying only the five basic cosmological parameters. It will be interesting to study how the new observable helps to lift degeneracies e.g. between the amplitude $A_s$ and the bias, which need a good resolution of redshift-space distortions. As has already been shown previously, the $\zeta$-variable is more sensitive to RSD than  galaxy number counts. 
%???%It will therefore also be interesting to study how it improves measurements of the growth factor $f(z)$.

Finally, the new window function $\om(z) = (\bar z -z)W(z)$ could also be replaced by a higher order polynomial in $z$. Constructing a series of orthogonal polynomials in this way, one can construct a series of new observables needing better and better redshift resolution. This can be considered as an alternative to ever slimmer redshift bins. However, as discussed in \cite{Matthewson:2021rmb}, very narrow redshift bins are affected by non-linearities even at very low multipoles. We expect the same to be true for high order polynomials of this new observable.
A detailed analysis of this is left to future work.

\section*{Acknowledgement}
We thank Louis Legrand for helpful discussions. This work is supported by  the Swiss National Science Foundation grant number 200020\underline{~~}182044.

%---------------------------------------------------------
\appendix
\section{Perturbation variables}\label{a:pert}
The density perturbation at fixed redshift is given by~\cite{Bonvin:2011bg}
\bea
\de_z(\bn,z) &=& b(z) D(\bn,z) -3\HH V(\bn,z)+3(\bV\cdot\bn)(\bn,z) +3\Psi(\bn,z)
\nonumber \\
&& \qquad + 3\int_{0}^{r(z)}(\dot\Psi+\dot\Phi)(\bn,z(r))dr \;.  \label{ea:dez2}
\eea 
$D\equiv  D_{cm} $ is the density fluctuation in co-moving gauge~\cite{Durrer:2020fza} and $b(z)$ is the galaxy bias. $V$ is the velocity potential so that $\bV=-\bnabla V$.

The volume perturbation is
\bea
\lefteqn{\frac{\de v}{v}=-2(\Psi+\Phi) -4\bV\cdot\bn +\frac{1}{\HH}
\left[\dot\Phi+\dd_r\Psi-\dd_r(\bV\cdot\bn)\right]} \nonumber\\
&&+\left(\frac{\dot{\HH}}{\HH^2}+\frac{2}{r(z)\HH}\right)
\left(\Psi+\bV\cdot\bn+ \int_{0}^{r(z)} dr(\dot{\Phi}+\dot{\Psi})\right)\nonumber\\
&&-3\int_{0}^{r(z)} dr(\dot{\Phi}+\dot{\Psi})+ \frac{2}{r(z)}\int_{0}^{r(z)} dr (\Phi+\Psi)\nonumber\\
&&- \frac{1}{r(z)}\int_{0}^{r(z)} dr\frac{r(z)-r}{r}
\Delta_\Om(\Phi+\Psi)~.  \label{ea:dev}
\eea
Here $\Delta_\Om$ denotes the Laplacian on the sphere.

Taking into account also magnification bias and evolution bias, as well as galaxy bias in comoving gauge, we obtain for the total number count fluctuation $\De$
\bea
\De(\bn,z,m_*) &=& b(z)D(L>\bar L_*)-3 {\cal H} V -(2-5s)\Phi + \Psi + \frac{1}{\HH}
\left[\dot\Phi+\dd_r(\bV\cdot\bn)\right] +  \nonumber \\  &&
 \left(\frac{{\dot\HH}}{\HH^2}+\frac{2-5s}{r(z)\HH} +5s-f_{\rm evo}\right)\left(\Psi+\bV\cdot\bn+ 
 \int_0^{r(z)}\hspace{-0.3mm}dr(\dot\Phi+\dot\Psi)\right) 
   \nonumber \\  &&  \label{DezNF}
+\frac{2-5s}{2r(z)}\int_0^{r(z)}\hspace{-0.3mm}dr \left[2-\frac{r(z)-r}{r}\Delta_\Om\right] (\Phi+\Psi) \,.
\eea
Here $m_*$ is the magnitude limit of the survey, $L_*$ is the corresponding luminosity limit and the evolution bias $f_{\rm evo}(z)$ is defined by
with
$$
f_{\rm evo} = \frac{\partial\ln\left(a^3 \bar N(z,L>\bar L_*)\right)}{\HH \partial \tau} \,.
$$
We have also introduced the magnification bias $s(z)$ via the  logarithmic derivative,
\be
s(z,m_*) \equiv \frac{\dd\log_{10} \bar N(z,m<m_*)}{\dd m_*} = \frac{\bar n_S(z,\bar L_*)}{2.5\bar N(z,L>\bar L_*)} \,,
\ee
where
\bea
\bar N(z,L>\bar L_*) \equiv \frac{\ln10}{2.5} \int_{-\infty}^{m_*} \bar n_S(z,m) dm  
= \int_{F_*}^{\infty} \bar n_S(z,\ln F) d\ln F \,.
\eea
In other words, $\bar n_S$ is the number density of sources per logarithmic flux interval.
Using this definition and the fact that at fixed $z$ logarithmic derivatives with respect to $L$ are the same as those with respect to $F$, we find
\be
\left. \frac{\partial \ln\bar n_S(z,\ln L)}{\partial\ln L} \right|_{\bar L_*} = -\frac{5}{2} s(z,m_*) \,,
\ee

%\den{I added the following as a refernce for the reader for what we mean when talking about density, RSD, lensing , and gravitational contributions:}
As it is done in the main text, the different terms in (\ref{DezNF}) can be divided into four principal categories: density, $b(z)D$, redshift-space distortions, $\HH^{-1}[\dd_r(\bV\cdot\bn) +(\dot\HH/\HH)\bV\cdot\bn]$, lensing proportional to $\int_0^{r(z)}\frac{r(z)-r}{r}\Delta_\Om (\Phi+\Psi)$, all remaining terms are referred to as large-scale gravitational (or relativistic) contributions.

\section{Flat sky approximation} \label{a:flat}
In this appendix we derive a useful approximation for $C_\ell^\zeta$ at equal redshift, $z_1=z_2$ for relatively slim bins.
For this we note that $\int (z-\bar z)\om(z,z_1)dz =0$ Therefore, if $\De(\bn,z)$ would not depend on redshift, $\zeta$ would vanish.  Expanding $\De$ to first order around $\bar z(z_1)$ we obtain
\be
\zeta(\bn,z_1) \simeq \frac{d}{dz}\De(\bn,\bar z(z_1))\,F_2(z_1) \quad \mbox{ where } \quad
F_2(z_1) =\int (z-\bar z(z_1))^2\om(z,z_1)dz \,.
\ee
$\De$ depends on redshift in two ways. On the one hand, the variable has to be evaluated at the position $\bx=r(z)\bn$ and on the other hand at conformal time $\tau(z)$. In terms of the variables $\bx$ and $\tau$ we find
\be
\left.\frac{d}{dz}\De\right|_{\bn} = \frac{1}{H}\left[\bn\cd\dd_\bx \left. \De\right|_\tau -\left. \dd_\tau\De\right|_\bx \right]
\ee
While the first term is typically of order $(k/H)\De$ for a mode with wave number $k$, the second term is typically of order $\De$. Therefore, on subhorizon scales the first term dominates. Let us concentrate on this dominant term here. For consistency we also only consider only the dominant terms in $\De$ which are density and redshift-space distortion (RSD),
\be
\De(\bn,z)\simeq b(z)D(r(z)\bn,\tau(z)) -\frac{1}{\HH}\dd_rV_r(r(z)\bn,\tau(z))\,,
\ee
where $D$ is the density fluctuation in comoving gauge and $\bV$ is the velocity perturbation in longitudinal gauge.
For the dominant $k/\HH$-term in the $z$-derivative we then obtain
\be
\dd_z\De(\bn,z)|_{\bn}\simeq \frac{1}{H}\Bigg[b(z)\dd_rD(r(z)\bn,t(z)) -\frac{1}{\HH}\dd^2_rV_r(r(z)\bn,t(z))\Bigg]\,,
\ee
This yields
\be
C_\ell^\zeta(z_1,z_1)= C_\ell^{\De'}(\bar z(z_1))F^2_2(z_1)%/(1+\bar z(z_1))^2
\ee
where $C_\ell^{\De'}$ is the power spectrum of $(1+z)\dd_z\De(\bn,z)|_{\bn}$.

In the flat sky approximation, see~\cite{Matthewson:2020rdt} for details, 
\bea
C_\ell^{\De'}(z) &=& \frac{1}{\pi[H(z)r(z)]^2}\int_0^\infty dk_\pa k_\pa^2\left(b(z)T_D(k,z) -\frac{k_\pa^2}{k \HH}T_V(k,z)\right)^2P_{\RR}(k) \,, \quad k=\sqrt{k_\pa^2+\frac{\ell^2}{r^2(z)}} \nonumber\\
 &=& \frac{1}{\pi [H(z)r(z)]^2}\int_0^\infty dk_\pa{k_\pa}^2\left[b(z)+f(z) \left(\frac{k_\pa}{k}\right)^2\right]^2T^2_D(k,z)P_{\RR}(k)\,.
\eea
where we have used
\be\label{e:TVTD}
T_V(k,z) = -\frac{\cal H}{k} f(z) T_D(k,z)\,.
\ee
Here $T_D$ and $T_V$ are the density and velocity transfer functions and $P_\RR(k)$ is the primordial curvature power spectrum, $(k^3/2\pi^2)P_\RR = A_s(k/k_*)^{n_s-1}$. Eq.~\eqref{e:TVTD} is a consequence of
the continuity equation which yields (at first order)  $-kV=\dot D=\HH f D$, where $f=d\log D/d\log a$ is the growth function.

Let us compare this expression with the one for the standard terms (density and RSD) in the galaxy number counts~\cite{Matthewson:2020rdt},
\be
C_\ell^{\De}(z) ~ = ~\frac{1}{\pi r(z)^2}\int_0^\infty dk_\pa \left[b(z)+f(z) \left(\frac{k_\pa}{k}\right)^2\right]^2T^2_D(k,z)P_{\RR}(k)\,.
\ee
The difference is simply a factor $(k_\pa/H)^2$ inside the $k_\pa$ integral.
Including a Gaussian window with standard deviation $\si_z$ in redshift, hence standard deviation $\si_z/H(z)$ in $r(z)$ we obtain, for small window sizes ($\si_z\lesssim 0.1$):
\bea
C_{\mbox{Gauss}}^{\De}(\ell,z) &=&\frac{1}{\pi r(z)^2}\int_0^\infty
 dk_\pa e^{-\frac{k^2_\pa\si^2_z}{H^2}}\left[b(z)+f(z)\left(\frac{k_\pa}{k}\right)^2\right]^2 T_D(k)^2   P_{\mathcal{R}}(k)   \,\\
 \mbox{and} && \nonumber\\
C_{\mbox{Gauss}}^{\De'}(\ell,z) &=&\frac{1}{\pi [H(z)r(z)]^2}\int_0^\infty
 dk_\pa k_\pa^2e^{-\frac{k^2_\pa\si^2_z}{H^2}}\left[b(z)+f(z)\left(\frac{k_\pa}{k}\right)^2\right]^2T_D(k)^2   P_{\mathcal{R}}(k)  \\
 &=& -\frac{\dd}{\dd\si_z^2}C_{\mbox{Gauss}}^{\De}(\ell,z)\,.
\eea
This shows also that the $\zeta$-spectrum is more sensitive than the number counts spectrum to high radial wave numbers.

\section{Shot Noise}
\label{sec:shotnoise}
Let us also derive a simple expression for the shot noise of $C_\ell^\zeta(z_i)=C_\ell^\zeta(z_i,z_i)$. We know that for different redshift bins $C_\ell^\zeta(z_i,z_j)$, $i\neq j$ there is no shot noise. Furthermore, the shot noise of $C_\ell^\De(z_i)$ is simply the inverse of the number density,
\be
C_\ell^{\De\,SN}(z_i) =\frac{4\pi f_{\rm sky}}{N_i}\,,
\label{eq:SNdelta}
\ee
where $N_i$ is the number of galaxies in the bin $i$ and $f_{\rm sky}$ is the observed sky fraction.
 The total measured $C_\ell^\De$ are given by $C_\ell^{\De\,{\rm (obs)}}(z_i) = C_\ell^{\De}(z_i) + C_\ell^{\De~SN}(z_i)$.
 We want to obtain this shot noise by an integration over the window function, $W(z_i,z)$. Knowing that the shot noise gives a constant contribution in each bin $i$ and that it comes from the correlation of each galaxy with itself, we can make the Ansatz $C_\ell^{\De\,SN}(z,z')= \al_i^{SN}\de(z-z')$, so that
 \bea
 C_\ell^{\De\,{\rm (obs)}}(z_i) &=&\int dzdz'W(z_i,z)W(z_i,z')[C^\De_\ell(z,z') +\al_i^{SN}\de(z-z')] \nonumber\\
 &=& C_\ell^{\De}(z_i) +\al_i^{SN}\int dz W^2(z_i,z) \\
 &=&  C_\ell^{\De}(z_i) 
  +\frac{4\pi f_{\rm sky}}{N_i}\,. \nonumber\\
 \eea
 This implies
 \be
 \al_i^{SN} =\frac{4\pi f_{\rm sky}}{N_i}\left[\int dz W^2(z_i,z)\right]^{-1} \,.
\ee
Note that even though the integral over $W$ is normalized, this is not the case for $W^2$ (except for a tophat window with width 1).
We now can insert this value of $\al_i^{SN}$ in expression \eqref{e:Clzeta} for $z_1=z_2=z_i$. With \eqref{e:deff} and \eqref{e:exzeta} this leads to 
\be
C_\ell^{\zeta\,{\rm (obs)}}(z_i)= C_\ell^{\zeta}(z_i) + C_\ell^{\zeta\,SN}(z_i)
\ee
with
\be
C_\ell^{\zeta\,SN}(z_i) = \frac{4\pi f_{\rm sky}}{N_i}
\left[\frac{\int dz (z-\bar z)^2\omega^2(z_i,z)}{\int dz W^2(z_i,z)}\right]\,.
\label{eq:SNzeta}
\ee

For the cross-correlations between ${\zeta}$ and $\Delta$, we find similarly:
\be
C_\ell^{\zeta\Delta\,SN}(z_i) = \frac{4\pi f_{\rm sky}}{N_i}
\left[\frac{\int dz (z-\bar z)\om(z_i,z)W(z_i,z)}{\int dz W^2(z_i,z)}\right]\,.
\label{eq:SNdeltazeta}
\ee
Since $\bar z$ is the mean of $z$ calculated with the window function $\om$, without the factor $W$ this integral would vanish.
Furthermore, as  $\om(z_i,z) \propto (r^2(z)/H(z))W(z_i,z)$, and within $\Lambda$CDM with standard parameters $r^2(z)/H(z)$ is a growing function of $z$ for $z\lesssim 2.4$, a mean with the window function $\om$ favours slightly larger values of $z$ than a mean with the window function $\om W$. Therefore the cross shot noise is not exactly vanishing but negative.%, see Fig~\ref{fig:shotnoise}.

\section{Fisher Analysis}
\label{sec:fisherappendix}
%\den{New Fisher analysis appendix with the $a_{lm}$'s as data}\\
In this appendix, we briefly review the formalism of how to obtain the best-fit cosmological parameters and their errors from observations, for a more detailed account we refer to \cite{Tegmark:1996bz,Albrecht:2009ct,Carron:2012pw} and references therein. The numerical results obtained in this paper use a simple python implementation of the mathematics presented here  which can be found a the URL \texttt{\url{https://github.com/WillMatt4/SMAL-FRY}}.
We consider a cosmological model determined by a set of $P$ parameters $\{\lambda_a\}_{a=1,...,P}$ and  observational data $\{d_i\}_{i=1,...,N}$. In the following text, we drop the curly brackets in favour of readability. The likelihood function $\mathcal{L}(\lambda_a)$ describes the probability of finding the measured data $d_i$ as function of the cosmological parameters $\lambda_a$, in terms of conditional probabilities: $\mathcal{L}(\lambda_a)= p(d_i|\lambda_a)$. For a Gaussian distribution of correlated measurements with correlation function $\mathcal{C}_{ij} = \braket{d_i d_j}$, the likelihood function reads
\begin{equation}
	\mathcal{L}(\lambda_a) = \frac{1}{\sqrt{(2\pi)^N \det \mathcal{C}}}\exp\left(-\frac{d_i \mathcal{C}_{ij}^{-1}(\lambda_a) d_j}{2}  \right)\,,
	\label{eq:likelihood}
\end{equation}
where $\mathcal{C}^{-1}$ is the matrix inverse of $\mathcal{C}$. In order to obtain the best-fit cosmological parameters $\lambda^*_a$, we need to maximise the probability for finding the observed data, i.e. the likelihood function, by varying the cosmological parameters: $\mathcal{L}(\lambda^*_a)\equiv \max\limits_{\lambda_a}\mathcal{L}(\lambda_a)$. For a Gaussian distribution, it is more convenient to maximise the logarithm of the likelihood, which, because of the monotonicity of the logarithm, yields the same result. In practice, finding the best-fit parameters is achieved for instance via a root finding routine, see e.g. \cite{Durrer:2020fza} for more details. In general, any likelihood function can be expanded within a sufficiently small neighbourhood of its maximum at $\lambda_a^*$ to yield a Gaussian:
\begin{align}
	\ln \mathcal{L}(\lambda_a) &= \ln \mathcal{L}(\lambda_a^*)+ \underbrace{\frac{\partial \ln \mathcal{L}}{\partial\lambda_a}\bigg|_{\lambda_a=\lambda_a^*}}_{=\,0} + \frac{1}{2}\frac{\partial^2 \ln\mathcal{L}}{\partial\lambda_a\partial\lambda_b}\bigg|_{\lambda_a=\lambda_a^*}\delta\lambda_a\,\delta\lambda_b+\mathcal{O}(\delta\lambda^3)\nonumber\\
	\Rightarrow \mathcal{L}(\lambda_a) &\simeq \mathcal{L}(\lambda_a^*)\exp\left(\frac{1}{2}\frac{\partial^2\ln\mathcal{L}}{\partial\lambda_a\partial\lambda_b}\delta\lambda_a\,\delta\lambda_b\right)\,,
	\label{eq:gaussexp}
\end{align}
with $\lambda_a=\lambda_a^*+\delta\lambda_a$ and implied summation over indices. The expectation value of the second order term is the Fisher matrix $F$:
\begin{equation}
	F_{ij}:=-\left\langle\frac{\partial^2\ln \mathcal{L}}{\partial \lambda_a\partial \lambda_b}\right\rangle\,,
\end{equation}
which can also directly be calculated from (\ref{eq:likelihood}) to yield
\begin{equation}
	F_{ij} = \frac{1}{2}\mathrm{tr} \left(\mathcal{C}^{-1}(\partial_i\mathcal{C}) \mathcal{C}^{-1}\partial_j\mathcal{C}\right)\,.
	\label{eq:fishermatrix}
\end{equation}
For a 1D Gaussian distribution $\mathcal{L}=\mathcal{L}_0\exp\left(-\frac{(x-\bar{x})^2}{2\sigma^2}\right)$, the second derivative is proportional to the variance: $\partial_x^2\ln\mathcal{L}\sim 1/\sigma^2$. Analogously, after a short calculation, see e.g. \cite{Durrer:2020fza}, one finds for the Gaussian regime (\ref{eq:gaussexp}) :
\begin{equation}
	\braket{\delta\lambda_i\delta\lambda_j} = F^{-1}_{ij}\,,
\end{equation}
hence, the error on the parameter $\lambda_a$, or more precisely, its marginalised error, is given by
\begin{equation}
	\delta\lambda_a = \sqrt{F^{-1}_{aa}}\,.
	\label{eq:error}
\end{equation}
Summarising, for given observational data $d_i$, we maximise the corresponding likelihood function $\mathcal{L}$ to find the best-fit cosmological parameters $\lambda_a$ reproducing the data. In order to arrive at their errors within the Gaussian approximation (\ref{eq:gaussexp}), we compute the Fisher matrix. The marginalised error $\delta\lambda_i$ of the i-th parameter is then given by the square root of the i-th diagonal element of the inverse Fisher matrix, cf. (\ref{eq:error}).

In the context of the present work, the data is given by the galaxy number counts, which can be decomposed into spherical harmonics:
\begin{equation}
	\Delta(z,\theta,\phi) = \sum_{\ell,m} = a_{\ell m}(z)\,Y(\theta,\phi)\quad.
\end{equation}
Therefore, we consider the coefficients $a_{\ell m}(z)$ as entries for the data vector $d$, which implies that the i-th entry is specified by the tuple ($\ell,m,z$). The covariance matrix is then given by
\begin{equation}
	\mathcal{C}_{ab} = \braket{d_ad_b} = \braket{a_{\ell m}(z_A)a_{\ell' m'}(z_B)}= C_\ell(z_A,z_B)\,\delta_{\ell\ell'}\delta_{m\,-m'}\quad,\label{eq:correlmatrix}
\end{equation}
where it was used that $a_{\ell m} = a^*_{\ell\,-m}$ due to $\Delta\in\mathbb{R}$. The indices $A,B$ run over the redshift bins. The inverse correlation matrix is then given by
\begin{equation}
	\mathcal{C}_{ab}^{-1} = [C_\ell(z_A,z_B)]^{-1}\,\delta_{\ell\ell'}\delta_{m\,-m'}\quad,\label{eq:inversecorrelmatrix}
\end{equation}
where $[\;]^{-1}$ denotes the matrix inverse. The Fisher matrix reads:
\begin{align}
	F_{ij} &= \frac{1}{2} \sum_{\substack{\ell,\ell',\ell'',\ell'''\\m,m',m'',m'''\\A,B,C,D}} [C_\ell(z_A,z_B)]^{-1}\delta_{\ell \ell'}\delta_{m\,-m'}\; \left(\partial_i C_{\ell'}(z_B,z_C)\right)\delta_{\ell'\ell''}\delta_{m'\,-m''}\;\times\nonumber\\
	&\hspace*{3cm}[C_{\ell''}(z_C,z_D)]^{-1}\delta_{\ell''\ell'''}\delta_{m''\,-m'''}\; \left(\partial_j C_{\ell'''}(z_D,z_A)\right)\delta_{\ell'''\ell}\delta_{m'''\,-m}\nonumber\\
	&= \frac{1}{2}\sum_{\ell,A,B,C,D} (2\ell+1)\;[C_\ell(z_A,z_B)]^{-1}\cdot\partial_iC_\ell(z_B,z_C)\cdot[C_\ell(z_C,z_D)]^{-1}\cdot\partial_jC_\ell(z_D,z_A)\,.
	\label{ea:FishAB}
\end{align}

If we are interested in forecasts based on more than one observable in the same redshift bins, the expression for the correlation matrix (\ref{eq:correlmatrix}) changes somewhat. For example, in the present paper, we consider the combined Fisher forecast based on both $\Delta$ and $\zeta$. In this case, let $i,j$ be labels for the observables $\Delta$ or $\zeta$. Then, instead of (\ref{eq:correlmatrix}) for a single observable, the expression for the correlation matrix now reads:
\begin{equation}
    \mathcal{C}_{ab} = \braket{a^i_{\ell m}(z_A)a^j_{\ell'm'}(z_B)} = C_\ell^{ij}(z_A,z_B)\,\delta_{\ell\ell'}\delta_{m\,-m'}\,. 
\end{equation}
For fixed $\ell$, $C_\ell^{ij}(z_A,z_B)$ is the matrix spanned by both observables in all redshift bins, as it can be seen in Figure \ref{fig:CorrCoeff}. It consists of two block matrices on the diagonal corresponding to $C_\ell^{\Delta\Delta}$ and $C_\ell^{\zeta\zeta}$ respectively, whereas the $\Delta\,\zeta$ correlations are stored in the off-diagonal part.
In the inverse in Eq.~\eqref{ea:FishAB} above, the inverse has to be taken not only wrt. the redshift bins, $z_A,z_B$ but also wrt. to the variables $i,j$. 
\ed{In our case the cross correlations $\langle\Delta\zeta\rangle$ are typically small and can be neglected when considering the auto-correlations $\langle\Delta\Delta\rangle$ and $\langle\zeta\zeta\rangle$.}
\vspace{2cm}

 \bibliographystyle{JHEP}
\bibliography{refs}

\providecommand{\href}[2]{#2}\begingroup\raggedright\begin{thebibliography}{10}

\bibitem{Hernandez-Monteagudo:2019epd}
C.~Hern\'andez\textendash{}Monteagudo, J.~Chaves-Montero, and R.~E. Angulo,
  {\it {Density weighted angular redshift fluctuations: a new cosmological
  observable}},  {\em Mon. Not. Roy. Astron. Soc.} {\bf 503} (2021), no.~1
  L56--L61, [\href{http://arxiv.org/abs/1911.12056}{{\tt arXiv:1911.12056}}].

\bibitem{Hernandez-Monteagudo:2020xnl}
C.~Hern\'andez-Monteagudo, J.~Chaves-Montero, R.~E. Angulo, and G.~Aric\`o,
  {\it {Tomographic constraints on gravity from angular redshift fluctuations
  in the late Universe}},  {\em Mon. Not. Roy. Astron. Soc.} {\bf 503} (2021),
  no.~1 L62--L66, [\href{http://arxiv.org/abs/2005.06568}{{\tt
  arXiv:2005.06568}}].

\bibitem{Legrand:2020sek}
L.~Legrand, C.~Hern\'andez-Monteagudo, M.~Douspis, N.~Aghanim, and R.~E.
  Angulo, {\it {High resolution tomography for galaxy spectroscopic surveys
  with Angular Redshift Fluctuations}},  {\em Astron. Astrophys.} {\bf 646}
  (2021) A109, [\href{http://arxiv.org/abs/2007.14412}{{\tt
  arXiv:2007.14412}}].

\bibitem{Lima-Hernandez:2022twl}
A.~Lima-Hern\'andez, C.~Hern\'andez-Monteagudo, and J.~Chaves-Montero, {\it
  {Relativistic Angular Redshift Fluctuations embedded in Large Scale Varying
  Gravitational Potentials}},  \href{http://arxiv.org/abs/2203.15008}{{\tt
  arXiv:2203.15008}}.

\bibitem{Bonvin:2011bg}
C.~Bonvin and R.~Durrer, {\it {What galaxy surveys really measure}},  {\em
  Phys. Rev.} {\bf D84} (2011) 063505,
  [\href{http://arxiv.org/abs/1105.5280}{{\tt arXiv:1105.5280}}].

\bibitem{Durrer:2020fza}
R.~Durrer, {\em {The Cosmic Microwave Background, 2nd Edition}}.
\newblock Cambridge University Press, 2020.

\bibitem{2009PhRvD..80h3514Y}
J.~{Yoo}, A.~L. {Fitzpatrick}, and M.~{Zaldarriaga}, {\it {New perspective on
  galaxy clustering as a cosmological probe: General relativistic effects}},
  {\em Phys. Rev. D} {\bf 80} (Oct., 2009) 083514,
  [\href{http://arxiv.org/abs/0907.0707}{{\tt arXiv:0907.0707}}].

\bibitem{Yoo:2010ni}
J.~Yoo, {\it {General Relativistic Description of the Observed Galaxy Power
  Spectrum: Do We Understand What We Measure?}},  {\em Phys.Rev.} {\bf D82}
  (2010) 083508, [\href{http://arxiv.org/abs/1009.3021}{{\tt
  arXiv:1009.3021}}].

\bibitem{Challinor:2011bk}
A.~Challinor and A.~Lewis, {\it {The linear power spectrum of observed source
  number counts}},  {\em Phys. Rev.} {\bf D84} (2011) 043516,
  [\href{http://arxiv.org/abs/1105.5292}{{\tt arXiv:1105.5292}}].

\bibitem{DiDio:2013bqa}
E.~Di~Dio, F.~Montanari, J.~Lesgourgues, and R.~Durrer, {\it {The CLASSgal code
  for Relativistic Cosmological Large Scale Structure}},  {\em JCAP} {\bf 11}
  (2013) 044, [\href{http://arxiv.org/abs/1307.1459}{{\tt arXiv:1307.1459}}].

\bibitem{CLASS}
J.~{Lesgourgues}, {\it {The Cosmic Linear Anisotropy Solving System (CLASS) I:
  Overview}},  {\em arXiv e-prints} (Apr., 2011) arXiv:1104.2932,
  [\href{http://arxiv.org/abs/1104.2932}{{\tt arXiv:1104.2932}}].

\bibitem{planck2018}
N.~Aghanim, Y.~Akrami, M.~Ashdown, J.~Aumont, C.~Baccigalupi, M.~Ballardini,
  A.~J. Banday, R.~B. Barreiro, N.~Bartolo, and et~al., {\it Planck 2018
  results},  {\em Astronomy \& Astrophysics} {\bf 641} (Sep, 2020) A6.

\bibitem{2020Euc}
A.~Blanchard, S.~Camera, C.~Carbone, V.~F. Cardone, S.~Casas, S.~Clesse,
  S.~Ilić, M.~Kilbinger, T.~Kitching, and et~al., {\it Euclid preparation},
  {\em Astronomy \& Astrophysics} {\bf 642} (Oct, 2020) A191.

\bibitem{2021SKA}
G.~Jelic-Cizmek, F.~Lepori, C.~Bonvin, and R.~Durrer, {\it On the importance of
  lensing for galaxy clustering in photometric and spectroscopic surveys},
  {\em Journal of Cosmology and Astroparticle Physics} {\bf 2021} (Apr, 2021)
  055.

\bibitem{2016SKA}
P.~Bull, {\it Extending cosmological tests of general relativity with the
  square kilometre array},  {\em The Astrophysical Journal} {\bf 817} (Jan,
  2016) 26.

\bibitem{2021Maartens}
R.~Maartens, J.~Fonseca, S.~Camera, S.~Jolicoeur, J.-A. Viljoen, and
  C.~Clarkson, {\it Magnification and evolution biases in large-scale structure
  surveys},  {\em Journal of Cosmology and Astroparticle Physics} {\bf 2021}
  (Dec, 2021) 009.

\bibitem{Matthewson:2020rdt}
W.~L. Matthewson and R.~Durrer, {\it {The Flat Sky Approximation to Galaxy
  Number Counts}},  {\em JCAP} {\bf 02} (2021) 027,
  [\href{http://arxiv.org/abs/2006.13525}{{\tt arXiv:2006.13525}}].

\bibitem{Matthewson:2021rmb}
W.~L. Matthewson and R.~Durrer, {\it {Small scale effects in the observable
  power spectrum at large angular scales}},
  \href{http://arxiv.org/abs/2107.00467}{{\tt arXiv:2107.00467}}.

\bibitem{Tegmark:1996bz}
M.~Tegmark, A.~Taylor, and A.~Heavens, {\it {Karhunen-Loeve eigenvalue problems
  in cosmology: How should we tackle large data sets?}},  {\em Astrophys. J.}
  {\bf 480} (1997) 22, [\href{http://arxiv.org/abs/astro-ph/9603021}{{\tt
  astro-ph/9603021}}].

\bibitem{Albrecht:2009ct}
A.~Albrecht et~al., {\it {Findings of the Joint Dark Energy Mission Figure of
  Merit Science Working Group}},  \href{http://arxiv.org/abs/0901.0721}{{\tt
  arXiv:0901.0721}}.

\bibitem{Carron:2012pw}
J.~Carron, {\it {On the assumption of Gaussianity for cosmological two-point
  statistics and parameter dependent covariance matrices}},  {\em Astron.
  Astrophys.} {\bf 551} (2013) A88, [\href{http://arxiv.org/abs/1204.4724}{{\tt
  arXiv:1204.4724}}].

\end{thebibliography}\endgroup

\end{document}